# Two-dimensional Quasi-Freestanding Molecular Crystals for High-Performance Organic Field-Effect Transistors

Daowei He,[1,†] Yuhan Zhang,[1,†] Qisheng Wu,[2] Rui Xu,[3] Haiyan Nan,[2] Junfang Liu,[3] Jianjun Yao,[4] Zilu Wang,[2] Shijun Yuan,[2] Yun Li,[1] Yi Shi,[1,*] Jinlan Wang,[2,*] Zhenhua Ni,[2] Lin He,[3] Feng Miao,[5] Fengqi Song,[5] Hangxun Xu,[6] K. Watanabe,[7] T. Taniguchi,[7] Jian-Bin Xu,[8] and Xinran Wang[1,*]

[1]*National Laboratory of Solid State Microstructures, School of Electronic Science and Engineering, and Collaborative Innovation Center of Advanced Microstructures, Nanjing University, Nanjing 210093, P. R. China*

[2]*Department of Physics, Southeast University, Nanjing 211189, P. R. China*

[3]*Department of Physics, Beijing Normal University, Beijing, 100875, P. R. China*

[4]*Asylum Research, Oxford Instruments, Shanghai, 200233, P. R. China*

[5]*School of Physics, Nanjing University, Nanjing 210093, P. R. China*

[6]*CAS Key Laboratory of Soft Matter Chemistry, Department of Polymer Science and Engineering, University of Science and Technology of China, Hefei 230026, P. R. China*

[7]*National Institute for Materials Science, 1-1 Namiki, Tsukuba, 305-0044, Japan*

[8]*Department of Electronic Engineering and Materials Science and Technology Research Center, The Chinese University of Hong Kong, Hong Kong SAR, P. R. China*

*\* Correspondence to X. W. (xrwang@nju.edu.cn), Y. S. (yshi@nju.edu.cn) or J. W. (jlwang@seu.edu.cn).*

† *These authors contribute equally to this work.*

**Abstract**

Two-dimensional atomic crystals are extensively studied in recent years due to their exciting physics and device applications. However, a molecular counterpart, with scalable processability and competitive device performance, is still challenging. Here, we demonstrate that high-quality few-layer dioctylbenzothienobenzothiophene molecular crystals can be grown on graphene or boron nitride substrate via van der Waals epitaxy, with precisely controlled thickness down to monolayer, large-area single crystal, low process temperature and patterning capability. The crystalline layers are atomically smooth and effectively decoupled from the substrate due to weak van der Waals interactions, affording a pristine interface for high-performance organic transistors. As a result, *monolayer* dioctylbenzothienobenzothiophene molecular crystal field-effect transistors on boron nitride show record-high carrier mobility up to $10 \text{cm}^2\text{V}^{-1}\text{s}^{-1}$ and aggressively scaled saturation voltage around 1V. Our work unveils an exciting new class of two-dimensional molecular materials for electronic and optoelectronic applications.

**Introduction**

The research on two-dimensional (2D) layered materials has been thriving for a decade now since the discovery of graphene[1-3]. Most of the 2D materials studied so far, including graphene, chalcogenides and oxides, are inorganic atomic crystals with strong in-plane chemical bonds[3]. 2D molecular crystals held by van der Waals (vdW) forces, on the other hand, are rarely reported. Organic molecular crystals represent an important class of materials for electronic and photonic applications[4-7]. Compared to bulk materials, monolayers could effectively eliminate interlayer screening, thus offer an ideal system to directly probe the effect of disorders[8-9] and interfaces[10-11] on charge transport. Furthermore, both carrier injection and modulation would become more efficient to greatly improve the performance of organic field-effect transistors (OFETs). Therefore, 2D molecular crystals can open a new paradigm in the applications of layered materials and heterostructures.

Here, we report on the epitaxial growth and OFETs of 2D dioctylbenzothienobenzothiophene ($C_8$-BTBT) molecular crystals down to monolayer on graphene and hexagonal boron nitride (abbreviated as BN) substrates. We select $C_8$-BTBT because it is among the highest mobility small-molecule organic materials[12-14]. Graphene and BN are also ideal substrates as they are atomically flat without dangling bonds[15]. Therefore, the initial stage of epitaxy relies on the vdW interaction between the molecules and substrate[16]. We show that the substrate flatness and weak vdW interaction are crucial for high-performance monolayer organic transistors because of the minimal disturbance in both crystal growth and charge transport.

**Results**

**vdW epitaxial growth and characterization of $C_8$-BTBT molecular crystals**

The growth was carried out in a tube furnace under high vacuum of ~$4\times10^{-6}$ Torr without carrier gas (Supplementary Figure 1). We heated up the $C_8$-BTBT powder (as received from NIPPON KAYAKU co., Ltd. without further purification) to 100-120 °C and placed graphene or BN on $SiO_2$/Si a few inches away from the source to start the growth. Below we focus our discussion on graphene substrate, while similar observations are also made on BN (Supplementary Figure 2 and 3). Fig. 1b-e show sequential atomic force microscopy (AFM) snapshots of the same sample during a 95-minute growth, which was intentionally interrupted for characterization (the sample was taken out to ambient each time, see Supplementary Figure 4 for the AFM images of the whole sequence). From these images, several conclusions can be drawn. i) The crystals preferentially grew on graphene, in a layer-by-layer fashion, with atomic smoothness. The thickness of the initial two layers (namely the interfacial layer, IL, and the first layer, 1L) was ~0.6nm and ~1.7nm respectively (Fig. 1f, Supplementary Figure 5), indicating that molecular packing in the initial layers was different from bulk crystals[12]. However, the thickness of the subsequent layers was ~3nm as bulk crystals. ii) In each layer, the growth initiated at certain nucleation sites, and proceeded nearly isotropically as compact islands. The most common nucleation sites were disorders from the previous layers (Fig. 1c) or from the substrate (Fig. 1d, Supplementary Figure 4a,4f and 4m), as well as edges (Fig. 1e, Supplementary Figure 4k,4n), likely due to their high surface energy. iii) The frequent interruption and ambient exposure of the sample did not significantly affect the growth, which suggested that the crystals were of pristine quality and stable against photo oxidation in ambient. Supplementary Figure 6 shows the growth process of another sample with excellent quality and thickness control.

The sample in Fig. 1b-e has multiple nucleation sites in each layer, probably due to the defects such as cracks and wrinkles of the underlying graphene (Supplementary Figure 4a).

However, with careful control of the growth parameters on defect-free graphene, we repeatedly observed large-area uniform 1L or 2L $C_8$-BTBT single crystals up to ~80μm in size (Supplementary Figure 7 - Supplementary Figure 11). Fig. 1g shows a uniform monolayer $C_8$-BTBT crystal grown on a graphene sample over 30μm long and 5μm wide (we do not count the IL in the number of layers because charge transport does not likely happen in this layer, as discussed later), as confirmed by the total thickness of 3.7nm (Fig. 1g inset) and Raman spectroscopy (Fig. 1h, Supplementary Figure 13b). Cross-polarized optical micrographs showed that the entire piece was a single crystal except a small portion at the bottom (Figs. 1i, j), and the optical intensity had a four-fold symmetry as expected for high-quality single crystal (Fig. 1k). The low nucleation density and island growth mode (as opposed to fractal growth) point to a high diffusion coefficient of $C_8$-BTBT molecules both on surface and along the island edges[17-18]. This is a natural result of the substrate flatness (Supplementary Figure 12) and relatively weak vdW forces from the substrate, in clear contrast to the conventional epitaxy where strong chemical bonds are formed at the interface[16].

The thickness of ~0.6nm for the IL pointed to a new form of molecular packing at the interface. To determine its structure, we performed combined scanning tunnelling microscopy (STM) and density functional theory (DFT) studies (see Methods for details). We observed under STM that the IL $C_8$-BTBT molecules were packed in a rectangular lattice with $d_1$=2.52nm and $d_2$=0.66nm in two orthogonal directions (Fig. 2a). Based on this morphology, we built a series of possible molecular configurations, where the $C_8$-BTBT molecule unit was leaning towards graphene by different angles compared to the "stand-up" configuration (Supplementary Figure 14). We found that the most stable single-molecule configuration had a leaning angle of ~35°, in which both alkyl chains and benzothiophene were in closest proximity and parallel to graphene (Supplementary Figure 14f). In such configuration, both *CH-π* interactions[21] and the *π-π* interactions[22] were maximized. When forming periodic crystals, however, the fully-relaxed benzothiophene plane was slightly tilted to form a 10° angle with graphene in DFT, due to inter-molecular interactions. We further calculated the stability as a function of lattice constants and found that $d_1$=2.47nm and $d_2$=0.64nm was the most stable configuration (Supplementary Figure 15), in excellent agreement with STM results. Moreover, the angle between the carbon chains and the benzothiophene was calculated to be 134°, very close to the measured angle of 140° by STM. Finally, quantitative match with both the layer thickness ~0.52nm and detailed STM height profiles (Supplementary Figure 23) allowed us to confirm the crystal structure of the IL drawn in Fig. 2b. Clearly, the symmetry of IL is different from that of the underlying graphene substrate because of the vdW nature of the epitaxy[16]. Our DFT calculations suggest that 2D growth of IL is energetically favourable due to the intermolecular interactions (Supplementary Figure 16-18, Supplementary note 1). In addition, a global energy minimum exists for different IL orientations with respect to the graphene substrate (Supplementary Figure 19, Supplementary note 2), making it possible to form large-area single-crystals as observed experimentally (Supplementary Figure 10). We note that IL does not likely contribute to lateral charge transport due to small inter-molecular π-π interactions.

Due to limited vertical conductivity, the structure of the subsequent layers was studied by high-resolution AFM. We found that the crystal structures for both 1L and 2L were monolithic with herringbone-type packing as in bulk crystals[12] (Fig. 2c, Fig. 2d, Supplementary Figure 9-11, Supplementary Figure 24). The lattice constants were $a$=6.24±0.25Å (6.69±0.13Å), $b$=8.31±0.05Å (7.89±0.25Å) for $C_8$-BTBT crystals on graphene (BN), each measured from 10 areas on different samples. The slight deviation from bulk crystal values was not understood, but could come from limited lateral resolution of AFM. We did not observe any statistical difference of lattice constants between 1L and 2L. The $C_8$-BTBT molecules in 1L seemed more inclined than bulk crystals[12]. Apparently, the role of substrate was much reduced in 1L and became negligible in 2L and above, because the vdW forces decayed rapidly as $r^{-6}$. Indeed, the calculated binding energy of 1L and graphene is 0.07eV per molecule (Supplementary Figure 20), much smaller than that of IL as well as intermolecular binding energy within 1L. Importantly, the charge transport in graphene is not significantly affected by the deposition of $C_8$-BTBT layers (Supplementary Figure 25, Supplementary Table 1). Considering the charge transport in organic transistors occurs near the semiconductor-dielectric interface[23-24], the few-layer $C_8$-BTBT crystals can thus be regarded as quasi-freestanding with minimal disturbance from the substrate.

For future device applications, it is important to demonstrate the growth of 2D molecular crystals on centimetre scale as well as patterning capability. We successfully grew large-area $C_8$-BTBT crystals on CVD graphene (~1cm in size, see Methods for CVD sample preparation), with over 90% coverage (Supplementary Figure 26-28, Supplementary note 4). AFM analysis indicated that the $C_8$-BTBT crystals on CVD graphene shared the same molecular packing as those on exfoliated graphene. However, due to the polycrystalline nature of the CVD graphene, and the presence of wrinkles, cracks and polymer residue, the $C_8$-BTBT films were polycrystalline with domain size around tens of micrometers (Supplementary Figure 28). Future works are needed to further improve the quality and uniformity of $C_8$-BTBT crystals on CVD graphene. Our approach also enables patterning of ultrathin organic crystals for device integration. Fig. 1m shows that $C_8$-BTBT crystal can be readily grown on graphene patterns etched by oxygen plasma, with excellent uniformity.

Cross-polarized optical micrographs verified that the whole area was a single crystal (Figs. 1n, 1o).

**FETs of 2D C$_8$-BTBT crystals**

Next we demonstrate several logic device applications of the epitaxial C$_8$-BTBT crystals. When grown on graphene, a vertical graphene-semiconductor heterostructure was readily obtained. We then deposited 100nm-thick Au layer as top electrode to form a vertical OFET (Fig. 3b inset). Compared with previously reported 2D layered heterostructures[25-27], our method is inherently scalable without the need for mechanical transfer. Fig. 3a and 3b show transfer ($J_{ds}$-$V_g$) and output ($J_{ds}$-$V_{ds}$) characteristics of a typical vertical FET comprising of 5 layers of C$_8$-BTBT, with a large on/off ratio ~1500 at room temperature. The device exhibited small but finite hysteresis (Supplementary Figure 30), likely due to the ambient molecules trapped at the C$_8$-BTBT/Au interface. The hysteresis could be minimized by optimizing the deposition process of the top electrode. Another device with thicker C$_8$-BTBT showed on/off ratio ~$10^6$ at room temperature (Supplementary Figure 31). The devices exhibited diode-like rectifying behavior similar to graphene barristors[28-29], indicating different conduction mechanisms for the two bias polarities. Considering the energy of the highest occupied molecular orbital (HOMO) and lowest unoccupied molecular orbital (LUMO) of C$_8$-BTBT was 5.39eV and 1.55eV, respectively[30], the band diagrams under various bias conditions were drawn in Fig. 3c. Under forward bias, the charge transport was thermionic emission as reflected by the Arrhenius plot in the high temperature regime (Fig. 3d). In this regime, holes were thermally activated over a Schottky barrier $\Phi_{SB}$ at the metal-semiconductor interface. $\Phi_{SB}$ was found to decrease linearly with $V_{ds}^{1/2}$ (Fig. 3d inset) due to the image force as in conventional Schottky junctions[31]. The extrapolated $\Phi_{SB}$ at zero $V_{ds}$ was

~140meV, consistent with the work function of Au. As we changed the Fermi energy of graphene by sweeping $V_g$, the barrier width was altered to realize transistor switching (Fig. 3c) as in the case of Schottky type carbon nanotube transistors[32]. The ultralow leakage current at the off state suggested the absence of pinholes in the crystal even on the graphene edges. Under reverse bias, the thermionic emission current became exponentially small, and the charge transport was dominated by tunnelling through the $C_8$-BTBT layers. This was confirmed by the current insensitivity to temperature (Fig. 3b, Supplementary Figure 31a-c) as expected for tunnelling mechanism[25]. The tunneling current is very small because the orientation of the $C_8$-BTBT molecules does not favour vertical charge transport. Therefore the vertical transistors are suitable for low-power applications.

We further demonstrated logic gates by integrating two vertical OFETs. For AND gate, we grew $C_8$-BTBT crystals on two adjacent graphene as inputs, and connected them using a common top electrode as output (Fig. 3e inset). Either diode under forward bias (with a low input) would be closed with a much lower resistance than the pull-up resistor, therefore create a low output. The gate showed excellent logic function, with the high and low output close to $V_{dd}$ and 0, respectively (Fig. 3e). For OR gate, two top electrodes as inputs were deposited on $C_8$-BTBT crystal grown on a single graphene as output (Fig. 3f inset). Either diode under forward bias (with a high input) would be closed with a much lower resistance than the pull-down resistor, therefore create a high output. The non-ideal output for (4, 0) and (0, 4) was due to the finite leakage current under reverse bias (Fig. 3f). More complex functionality can be realized with lithographic patterning.

Finally, we fabricate planar OFETs with 1L $C_8$-BTBT crystals grown on insulating BN (Fig. 4, Supplementary Figure 32, 33). Despite the monolayer thickness (~1.7nm), the devices exhibited several features of ideal OFETs: linear $I_{ds}$-$V_g$ ($I_{ds}^{1/2}$-$V_g$) characteristics in the linear (saturation) regime, linear $I_{ds}$-$V_{ds}$ characteristics at low bias, and negligible hysteresis[4]. The room-temperature peak field-effect mobility $\mu$ could reach up to ~10cm$^2$V$^{-1}$s$^{-1}$ (Fig. 4c), much higher than previously reported values for monolayer OFETs (in the range of ~10$^{-6}$-0.1 cm$^2$V$^{-1}$s$^{-1}$)[33-37]. Note that $\mu$ was still underestimated due to contact resistance. Interestingly, $\mu$ was found to decrease after reaching a plateau as the carrier density gradually increased (Fig.4c, Supplementary 33a) due to additional sources of scattering such as electron-electron interactions[38]. Such drastic improvement indicated that the density of charge traps and grain boundaries was significantly reduced[8]. This was attributed to the pristine, ultrasmooth crystals and their weak coupling with the substrate, analogous to the case of graphene on BN[15]. The mobility was found to decrease slightly at low temperature in some of our devices (Supplementary Figure 33c). The weak insulating behavior pointed to finite density of charge traps[9,11], which will be carefully analyzed and improved in future works. The performance of monolayer $C_8$-BTBT FETs was comparable to 2D atomic crystals such as MoS$_2$ [39-42], but with a clear advantage of much lower synthesis temperature.

Compared to bulk crystal devices[12-14,43], the $I_{ds}$-$V_{ds}$ characteristics in Fig. 4b exhibited two unique features: the absence of non-linearity at low bias and an complete saturation with extremely small saturation voltage (~1V). Both features were direct consequence of the monolayer nature of the channel. The former appeared because the source/drain electrodes were in direct contact with the charge transport layer, leading to extremely efficient carrier injection. The latter was due to easy and complete channel pinch-off without interlayer screening effects[44]. With scaled dielectric thickness and other optimized device parameters, the operation voltage of these OFETs can be reduced to 1V, which makes these devices compatible with silicon CMOS in circuit applications.

**Discussion**

In conclusion, we demonstrate that vdW epitaxy on 2D materials provides a viable approach to grow few-layer organic molecular crystals with pristine quality, monolayer thickness control and large-area processability for high-performance and low-cost logic device applications. Compared to the previous efforts to grow organic crystals on 2D materials[29,43,45], our results truly demonstrate the advantage of this concept by proving the high quality of the interface.

**Methods**

**Sample preparation and growth process of $C_8$-BTBT crystals on graphene and BN**

We exfoliated graphene and BN on 285nm $SiO_2$/Si as growth substrate without further thermal treatment. The graphene substrates used in this work were monolayer unless otherwise stated. The graphene and BN were characterized by optical microscope, AFM and Raman spectroscopy before growth to obtain the thickness and topological information. For CVD graphene, we first grew the graphene on copper foils (Alfa Aesar) at 1050 °C under 4sccm $CH_4$ and 200sccm $H_2$. With poly(methyl methacrylate) (PMMA) as a support layer, we transferred a 1cm x 1cm CVD graphene on to $SiO_2$/Si substrate, followed by a thermal annealing at 300 °C in forming gas. The growth of $C_8$-BTBT crystals was carried out in a home-built tube furnace. We put the source and sample in the quartz tube and used a turbo molecular pump to evacuate the quartz tube to ~$4\times10^{-6}$Torr (Supplementary Figure 1). The distance between source and sample was accurately measured each time to attain good reproducibility. We then heated up the $C_8$-BTBT crystal to 100-120 °C to start the growth. In our experiment, the growth of $C_8$-BTBT layers was mainly controlled by the source temperature and growth time. To achieve precise control of IL, 1L and 2L growth (e.g. the samples in Fig. 1 and Supplementary Figure 5-8), the source temperature was 100 °C. The growth time to achieve uniform IL and 1L was approximately 15 minutes and 30 minutes, respectively, but was sample dependent. For thicker $C_8$-BTBT layers (e.g. the samples in Fig. 3), the source temperature was 120 °C. To terminate growth, we turned off the furnace and let the sample cool down to room temperature under high vacuum.

**AFM, Raman spectroscopy, cross-polarized optical microscopy and STM characterizations of the $C_8$-BTBT crystals**

Two types of AFM were performed in this work. For high-resolution AFM, the experiments were performed on an Asylum Cypher under ambient conditions with Asylum ARROW UHF AFM tips. In order to get sub-nanometre resolution, samples had to be ultra-flat. So we grew $C_8$-BTBT crystals on thick graphite and BN for high-resolution AFM. In this case, the roughness of the substrate did not affect the sample. We did not observe any difference in $C_8$-BTBT crystals grown on thick graphite and graphene in terms of the thickness and growth mode of each layer. For regular AFM, the experiments were performed on a Veeco Multimode 8 under ambient conditions.

Raman spectroscopy and mapping were performed on a WITec Alpha 300R confocal Raman system with a 532nm laser excitation (spot size ~300nm, laser power 1mW).

Cross-polarized optical microscopy was performed on the same WITec Alpha 300R Raman microscope with white light illumination and 50× objective. During the experiments, the laser notch filter was taken out and two cross-polarized linear polarizers were installed, one between illumination source and sample, the other between sample and detector. The images were acquired by scanning the sample. We integrated the spectrum from 520nm to 550nm to plot the images.

The STM study was carried out in a UNISOKU ultrahigh vacuum four-probe SPM. All STM measurements were performed at liquid-nitrogen temperature and the images were taken in a constant-current scanning mode. The STM tips were obtained by chemical etching from a wire of Pt (80%) Ir (20%) alloys. Lateral dimensions observed in the STM images were calibrated using a standard graphene lattice. For better substrate conductivity, the STM samples were prepared on monolayer CVD graphene grown on Cu foils. The samples were overall very flat with small corrugated areas and with some steps on the copper surface (Supplementary Figure 21, 22). Since high resolution STM scans were usually recorded in small areas ~100nm$^2$, the substrate roughness was very small and does not affect the STM measurement.

**Details of DFT calculations**

The DFT calculations were carried out within the framework of plane-wave density functional theory, implemented in the Vienna ab initio simulation package (VASP)[46]. We employed the projector-augmented-wave potentials[47] to describe the electron-ion interaction and the Perdew-Burke-Ernzerhof generalized gradient approximation (PBE-GGA) for exchange-correlation functional[48]. The effect of vdW interactions was described using the semiempirical correction scheme of Grimme, DFT-D2[49]. The kinetic energy cutoff of 400eV was adopted for the plane-wave expansion and the Brillouin zone was sampled by the Monkhorst-Pack scheme. All atomic positions were fully optimized until the maximum Hellmann-Feynman forces acting on each atom was less than 0.02 eVÅ$^{-1}$.

**Fabrication process of OFETs**

For the vertical OFETs in Fig. 3, we started by exfoliating monolayer graphene on 285nm SiO$_2$/Si. We transferred a 100nm-thick Au electrode to contact part of the graphene as probing pad before growth of C$_8$-BTBT crystals. Briefly, we pre-deposited 100nm Au film on a glass slide. Then we used a tungsten probe tip attached to a micro manipulator to carefully pick up the Au film and transfer to the target location under microscope[50]. AFM was performed before and after growth to extract the thickness of the crystals. After growth of the crystal, another Au film was transferred as top electrode, which partly overlapped with graphene (but not with the bottom Au electrode). We used mechanical transfer to deposit the Au electrodes to avoid any residue introduced by lithography and any damage of the ultrathin $C_8$-BTBT layers by vacuum deposition. Other than the growth, all other fabrication processes were carried out under ambient condition.

For the vertical OR gate, the fabrication was similar except that after growth, two separate Au films were used as inputs. The lateral separation between the Au electrodes (tens of micrometers) was three orders of magnitude larger than the separation between Au and bottom graphene (tens of nanometers), therefore the lateral charge transport was negligible. For the vertical AND gate, two adjacent monolayer graphene samples were identified after exfoliation. We transferred two Au electrodes, each contacting one of the graphene samples as probing pads. After growth, an Au film was transferred to overlap both graphene as common top electrode (but not with the bottom Au electrodes).

For the planar OFET in Fig. 4, we first exfoliated BN (with thickness normally smaller than 10nm) on 285nm $SiO_2$/Si substrate. The growth of monolayer $C_8$-BTBT crystal was carefully done by heating the source to 120 ℃. AFM was performed before and after growth to extract the thickness of the crystals. Finally, two Au films were transferred on the top of the $C_8$-BTBT crystal as source and drain electrodes.

**Electrical measurements of FETs**

Electrical measurements were carried out by an Agilent B1500 semiconductor parameter analyzer in a close-cycle cryogenic probe station with base pressure ~$10^{-5}$ Torr. The vertical transistors were annealed in vacuum at 120 °C before measurement to improve contacts, while the planar transistors did not need any annealing step.

**Acknowledgements**

The authors thank NIPPON KAYAKU co., Ltd. Japan for providing $C_8$-BTBT materials. We thank Hongjie Dai and Zhenan Bao for helpful discussion, and Yunqi Liu and Wenping Hu for assistance in device fabrication. This work was supported in part by Chinese National Key Fundamental Research Project 2013CBA01604, 2010CB923401, 2011CB302004; National Natural Science Foundation of China 61325020, 61261160499, 11274154, 21173040, 21373045, 61306021, 61229401; National Science and Technology Major Project 2011ZX02707, Natural Science Foundation of Jiangsu Province BK2012302, BK20130016, BK20130579; Specialized Research Fund for the Doctoral Program of Higher Education 20120091110028, and Research Grant Council of Hong Kong SAR N_CUHK405/12. J. W. acknowledge the computational resource at SEU and National Supercomputing Center in Tianjin for DFT calculations.

**Author contributions**

X. W. and Y. S. conceived and supervised the project. D. H., Y. Z., R. X., H. N., J. L., Y. L. and J. Y. performed experiments and data analysis. Q. W., Z. W., S. Y. and J. W. performed DFT calculations. Y. L., L. H., Z. N., F. M., F. S., H. X. and J.-B. X. contributed to data analysis. K. W. and T. T. provided BN samples. X. W., Y. S., J. W., L. H. and D. H.

co-wrote the paper. All authors discussed the results and commented on the manuscript. Correspondence and requests for materials should be addressed to X. W. or Y. S. for general aspects of the paper and to J. W. for DFT calculations.

**Competing financial interests**

The authors declare that they have no competing financial interests.

1818

**FIGURE CAPTIONS**

**Fig. 1.** Epitaxial growth of $C_8$-BTBT molecular crystals on graphene. (a) Cartoon illustration of the molecular structure of $C_8$-BTBT (left panel) and their packing on graphene (right panel). (b)-(e) Sequential AFM snapshots of a sample at different stages during a 95-minute growth. The scale bars are 2μm. The complete sequence of the growth process is show in Supplementary Figure 4. (f) Histogram of layer thickness of $C_8$-BTBT molecular crystals on graphene, taken from over 10 samples. (g)-(j) AFM (g), Raman mapping (h) and Cross-polarized optical micrographs (i, j) of a uniform monolayer $C_8$-BTBT crystal grown on graphene. The scale bars are 3μm. Inset of (g) is the height profile along the dashed line. The total thickness of graphene and the crystal is 3.7nm, confirming monolayer $C_8$-BTBT (2L would have made the total thickness greater than 5.3nm). The Raman spectrum of the $C_8$-BTBT crystal is plotted in Supplementary Figure 13b. (k) Normalized intensity of the monolayer $C_8$-BTBT crystal under cross-polarized optical microscope as a function of rotation angle. The data are taken at the marked spot in (i) and (j). (l)-(o) Patterned growth of $C_8$-BTBT crystal on graphene. The scale bars are 7μm. (l), (m) Optical microscopy image of a plasma-patterned graphene before and after $C_8$-BTBT growth, respectively. (n), (o) Cross-polarized optical micrographs of the same area after $C_8$-BTBT growth. The uniform color change over the entire area confirms that $C_8$-BTBT forms a single crystal.

**Fig. 2.** Molecular structure of the $C_8$-BTBT crystals on graphene. (a) A constant-current STM image of the IL on CVD graphene ($V_{sample}$=-0.91V and $I$=12.9pA). Scale bar, 2nm. (b) Top view (top panel) and side view (bottom panels) of the most stable IL structure obtained by DFT calculations. The lattice constants of 2.47nm and 0.64nm are in excellent agreement with experiments. (c) High-resolution AFM image of the 2L on graphene. The unit cell is marked. Inset is the Fast Fourier Transform of the AFM image with lattice indices. Scale bar, 1nm. (d) Top view (left panel) and side view (right panel) of the 2L structure according to the AFM image in (c).

**Fig. 3.** Vertical OFETs based on graphene-$C_8$-BTBT heterostructures. (a) Room temperature $J_{ds}$-$V_g$ characteristics of a device with ~15nm thick (5-layer) $C_8$-BTBT. From top to bottom, $V_{ds}$=2V and 1V, respectively. (b) Room temperature $J_{ds}$-$V_{ds}$ characteristics of the same device in (a). From top to bottom, $V_g$=-100V, -90V, -80V, -70V and 0V, respectively. Inset shows the device schematics. The source (S), drain (D) and gate (G) terminals are marked. (c) Energy band diagrams of the vertical OFET under various bias conditions. The band diagram under $V_{ds}$=0 and $V_g$=0 is plotted in Supplementary Figure 29. (d) Arrhenius plot (symbols) of normalized $J_{ds}$ of the same device in (a) under $V_g$=-100V. From top to bottom, $V_{ds}$=2V, 1.6V, 1.2V and 0.8V, respectively. The plot clearly shows two regimes: thermionic emission at high temperatures, and tunneling at low temperatures. Lines are theoretical fittings of the thermionic emission current. The extracted $\Phi_{SB}$ are plotted in the inset as a function of $V_{ds}^{1/2}$ (symbols), with a linear fitting (line). (e) (f) Output voltage levels of an AND (e) and OR (f) gate comprised of two vertical diodes. The insets show the circuit diagrams. The pull-up resistor for the AND gate and the pull-down resistor for the OR gate are both $5\times10^9$ Ohms. During the operation of both logic gates, $V_g$=-100V, $V_{dd}$=4V.

**Fig. 4.** Planar OFET of monolayer $C_8$-BTBT molecular crystals on BN. (a) Room temperature double-sweep $I_{ds}$-$V_g$ characteristics ($V_{ds}$=-0.5V), with little hysteresis. Black and blue lines are drawn in linear and log scales respectively. (b) $I_{ds}$-$V_{ds}$ characteristics of the device in (a). From top to bottom, $V_g$=-10V, -25V, -30V, -35V and -40V, respectively. Inset shows the optical microscopy image of the device. Scale bar, 16μm. (c) The extracted μ-$V_g$ relationship at room temperature. The peak mobility is over $10 cm^2V^{-1}s^{-1}$ for this device

**References**

1. Novoselov, K. S. & Geim, A. K. The rise of graphene. *Nature Mater.* **6**, 183–191 (2007).

2. Wang, Q., Kalantar-Zadeh, K., Kis, A., Coleman, J. N., & Strano, M. S. Electronics and optoelectronics of 2D transision metal dichalcogenides. *Nature Nanotech*. **7**, 699-712 (2012).

3. Geim, A. K. & Grigorieva, I. V. Van der Waals heterostructures. *Nature* **499**, 419-425 (2013).

4. Podzorov, V. & Guest. Organic single crystal: Addressing the fundamentals of organic electrnics. *MRS Bulletin* **38**, 15-24 (2013).

5. Sirringhaus, H. Device Physics of Solution-Processed Organic Field-Effect Transistors. *Adv. Mater.* **17**, 2411–2425 (2005).

6. Sirringhaus, H., Tessler, N. & Friend, R. H. Integrated optoelectronic devices based on conjugated polymers. *Science* **280**, 1741-1744 (1998).

7. Rogers, J. A. Toward Paperlike Displays. Science 291, 1502-1503 (2001).

8. Tello, M., Chiesa, M., Duffy, C. M. & Sirringhaus, H. Charge Trapping in Intergrain Regions of Pentacene Thin Film Transistors. *Adv. Funct. Mater.* **18**, 3907–3913 (2008).

9. Lezama, I. G. & Morpurgo, A. F. Progress in organic single-crystal field-effect transistors. *MRS Bulletin* **38**, 51-56 (2013).

10. Chua, L-L., Zaumseil, J., Chang, J-F., Sirringhaus, H., Friend, R. D. et al. General observation of n-type field-effect behavior in organic semiconductors. *Nature* **434**, 194-199 (2005).

11. Podzorov,V., Menard, E., Borissov, A., Kiryukhin, V., Rogers, J. A. & Gershenson, M. E. Intrinsic charge transport on the surface of organic semiconductors. *Phys. Rev. Lett.* **93**, 086602 (2004).

12. Minemawari, H., Yamada, T., Matsui, H., Chiba, R., Kumai, R. et al. Inkjet printing of single-crystal films. *Nature* **475**, 36-367 (2011).

13. Ebata, H., Izawa, T., Miyazaki, E., M., Kuwabara, H., Tatsuto Yui, T. et al. Highly Soluble [1]Benzothieno[3,2-b]benzothiophene (BTBT) Derivatives for High-Performance, Solution-Processed Organic Field-Effect Transistors. *J. Am. Chem. Soc.* **129**, 15732-15733 (2007).

14. Yuan, Y., Giri, G., Ayzner, A. L., Zoombelt, A. P., Huang, J., Bao, Z. et al. Ultra-high mobility transparent organic thin film transistors via off-center spin coating method. *Nature Commun*. **5**, 3005 (2014).

15. Dean, C. R., Young, A. F., Meric, I., Lee, C., Wang, L., Hone, J. et al. Boron nitride substrates for high quality graphene electronics. *Nature Nanotech.* **5**, 722-726 (2010).

16. Koma, A., Van der Waals epitaxy: a new epitaxial growth method for a highly lattice-mismatched system. *Thin Solid Films.* **216** , 72-76 (1992).

17. Zhang, Z., & Lagally, M. G. Atomistic processes in the early stages of thin-film growth. *Science* **276**, 377-383 (1997).

18. Meyer zu Heringdorf, F-J. Reuter, M. C., & Tromp, R. M. Growth dynamics of pentacene thin films. *Nature* **412**, 517-520 (2001).

21. Tsuzuki, S. CH/π Interactions. *Annu. Rep. Prog. Chem., Sect. C: Phys. Chem.* **108**, 69–95 (2012).

22. Hunter, C. A. & Sanders, J. K. M. The Nature of π-π Interactions. *J. Am. Chem. Soc.* **122**, 11450-11458 (2000).

23. Dodabalapur, A., Torsi, L. & Katz, H. E. Organic transistors: two-dimensional transport and improved electrical characteristics. *Science* **268**, 270-271 (1995).

24. Podzorov, V., Menard, E., Borissov, A., Kiryukhin, V., Rogers, J. A. & Gershenson, M. E. Intrinsic charge transport on the surface of organic semiconductors. *Phys. Rev. Lett*. **93**, 086602 (2004).

25. Britnell, L., Gorbachev, R. V., Jalil, R., Geim, A. K., Novoselov, K. S., Ponomarenko, L. A. et al. Field effect tunnelling transistors based on vertical graphene heterojunctions. Science 24, 947-950 (2012).

26. Georgiou, T., Jalil, R.,Belle, B. D., Geim, A. K., Novoselov, K. S., Mishchenko, A. et al. Vertical FET based on graphene-WS2 heterostructures for flexible and transparent electronics;. *Nature Nanotech*. **8**, 100-103 (2013).

27. Yu, W., Li, Z., Zhou, H., Chen, Y., Wang, Y., Duan, X. et al. Vertically stacked multi-heterostructures of layered materials for logic transistors and complementary inverters. *Nature Mater.* **12**, 246-252 (2013).

28. Yang, H., Heo, J., Park, S., Song, H. J., Seo, D. H., Kim, K. et al. Graphene barristor, a triode device with a gate controlled Schottky barrier. *Science* **336**, 1140-1143(2012).

29. Ojeda-Aristizabal, C., Bao, W. & Fuhrer, M. S. Thin-film barristor, a gate-tunable vertical graphene-pentacene device. *Phys. Rev. B* **88**, 035435 (2013).

30. Kobayashi, H., Kobayashi, N., Hosoi, S., Koshitani, N., Tokita, Y., & Itabashi, M. Hopping and band mobilities of pentacene, rubrene, and 2,7-

dioctyl[1]benzothieno[3,2-b][1]benzothiophene (C$_8$-BTBT) from first principle. *J. Chem. Phys. 139*, 014707 (2013).

31. Sze, S.M, & Ng, K. K. Physics of semiconductor devices, Wiley, 2006.

32. Martel, R., Derycke, V., Lavoie, C., Appenzeller, J., Chan, K. K., Tersoff, J. & Avouris, Ph. Ambipolar electrical transport in semiconducting single-wall carbon nanotubes. *Phy s. Rev. Lett.* **87**, 256805 (2001).

33. Smits, E. C. P., Mathijssen, S. G. J., Hal, P. A. van, Setayesh, S., Geuns, T. C. T., de Leeuw, D. M. et al. Bottom-up organic integrated circuits. *Nature* **455**, 956-959 (2008).

34. Mathijssen, S. G. S., Smits, E. C. P., van Hal, P. A., Wondergem, H. J., de Leeuw, D. M. et al. Monolayer coverage and channel length set the mobility in self-assembled monolayer field-effect transistors. *Nature Nanotech.* **4**, 674-680 (2009).

35. Li, L., Gao, P., Wang, W., Müllen, K., Fuchs, H. and Chi, L. Growth of Ultrathin Organic Semiconductor Microstripes with Thickness Control in the Monolayer Precision. *Angew. Chem. Int. Ed.* **52**, 12530-12535 (2013).

36. Ruiz, R., Papadimitratos, A., Mayer, A. C. and Malliaras, G. G. Thickness Dependence of Mobility in Pentacene Thin-Film Transistors. *Adv. Mater.*, **17**, 1795-1798 (2005).

37. Dinelli, F., Murgia, M., Levy, P., Cavallini, M., Biscarini, F. & de Leeuw, D. M. Spatially Correlated Charge Transport in Organic Thin Film Transistors. *Phys. Rev. Lett.* **93**, 116802 (2004).

38. Fratini, S., Morpurgo, A. F. & Ciuchi, S. Electron-phonon and electron-electron interactions in organic field effect transistors. *J. Phys. Chem. Sol.* **5**, 2195-2198 (2008).

39. Qiu, H., Xu, T., Wang, Z., Ren, W., Wang J., Wang, X. et al. Hopping transport through defect-induced localized states in MoS2. *Nature Commun.* **4**, 2642 (2013).

40. Qiu, H., Pan, L., Yao, Z., Li, J., Shi, Y., Wang, X. et al. Electrical characterization of back-gated bi-layer MoS2 field-effect transistors and the effect of ambient on their performances;. *Appl. Phys. Lett.* **100**, 123104 (2012).

41. Zhu, W., Low, T., Lee, Y-H., Wang, H., Xia, F., Avouris, P. et al. Electronic transport and device prospects of monolayer MoS2 disulfide grown by CVD. *Nature Commun.* **5**, 3087(2014).

42. Liu, H., Si, M., Najmaei, S., Neal, A. T., Lou, J., Ye, P. D. et al. Statistical study of deep submicron dual-gated FET on monolayer CVD MoS2. *Nano Lett.* **13**, 2640-2646 (2013).

43. Lee, C.-H. et al. Epitaxial Growth of Molecular Crystals on van der Waals Substrates for High-Performance Organic Electronics. *Adv. Mater.*, **26**, 2812- 2817(2014).

44. Meric, I., Han, M. Y., Young, A. F., Ozyilmaz, B., Kim, P., & Shepard, K. L. Current saturation in zero-bandgap, top-gated graphene FETs. *Nature Nanotech.* **3**, 654-659 (2008).

45. Lee, W. H., Park, J., Sim, S. H., Lim, S., Kim, K. S., Cho, K. Surface directed molecular assembly of pentacene on monolayer graphene for high performance organic transistors. *J. Am. Chem. Soc.***133**, 4447-4454 (2011).

46. Kresse, G. & Furthmüller, J. Efficient iterative schemes for *ab* initio total-energy calculations using a plane-wave basis set. *Phys. Rev. B* **54**, 11169-11186 (1996).

47. Blöchl, P. E. Projector augmented-wave method. *Phys. Rev. B* **50**, 17953-17979 (1994).

48. Perdew, J. P., Burke, K. & Ernzerhof, M. Generalized Gradient Approximation Made Simple. *Phy.s Rev. Lett.* **77**, 3865-3868 (1996)

49. Grimme, S. Semiempirical GGA-type density functional constructed with a long-range dispersion correction. *J. Comp. Chem.* **27**, 1787-1799 (2006).

50. Wang, X., Xu, J.-B., Wang, C., Du, J. & Xie, W. High-performance graphene devices on $SiO_2$/Si substrate modified by highly ordered self-assembled monolayers. *Adv. Mater.* **23**, 2464-2468 (2011).

Figure 1:

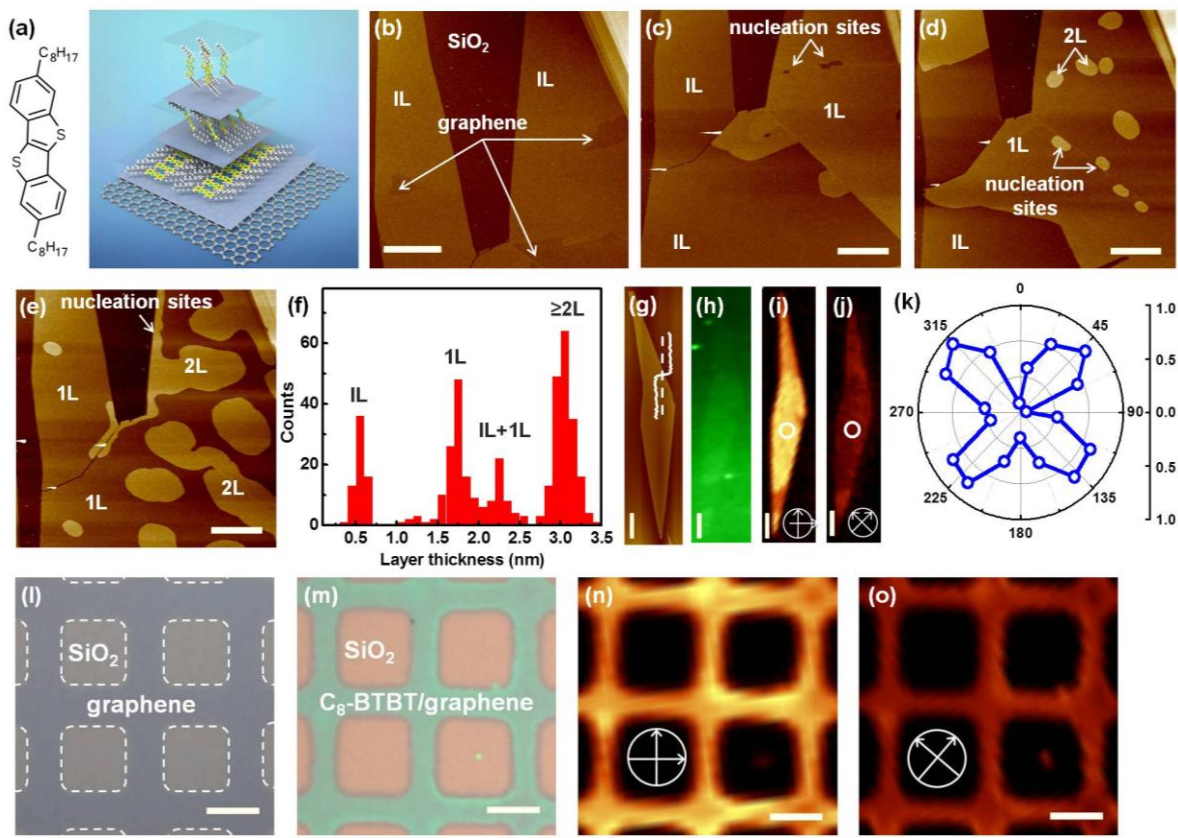

Figure 2:

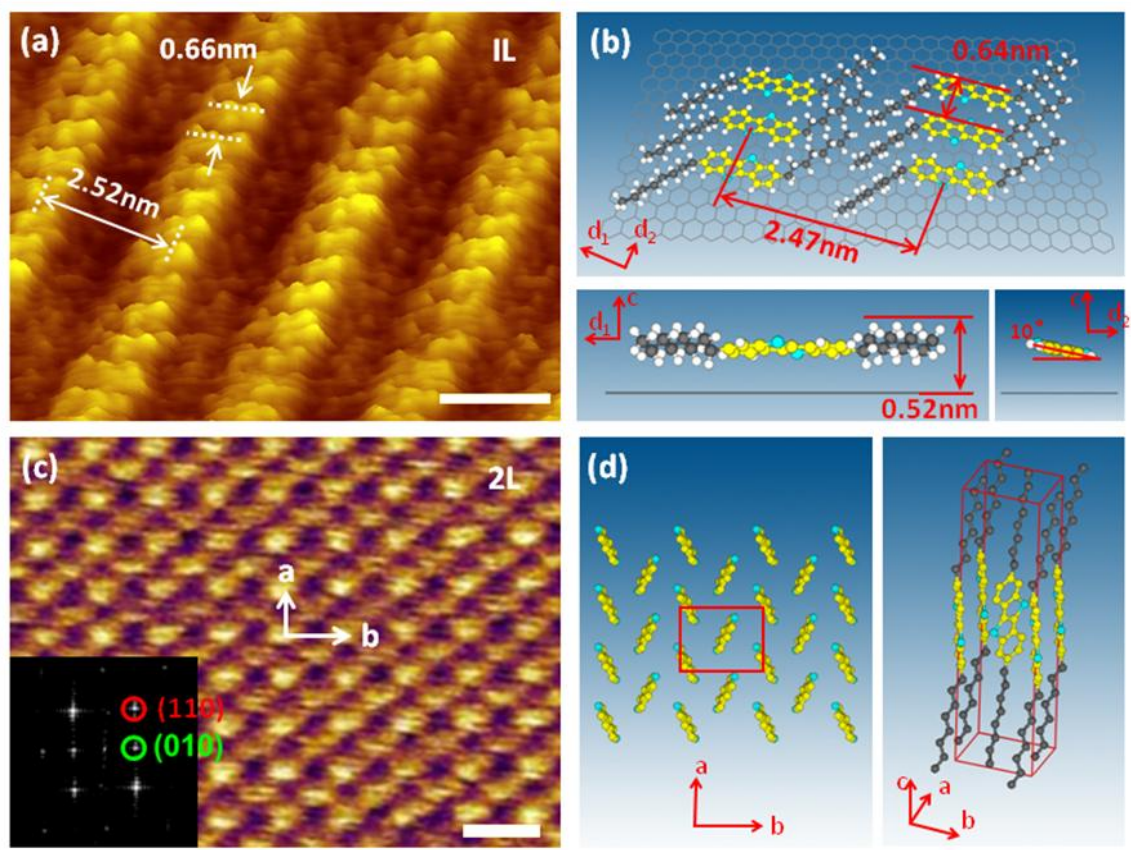

Figure 3:

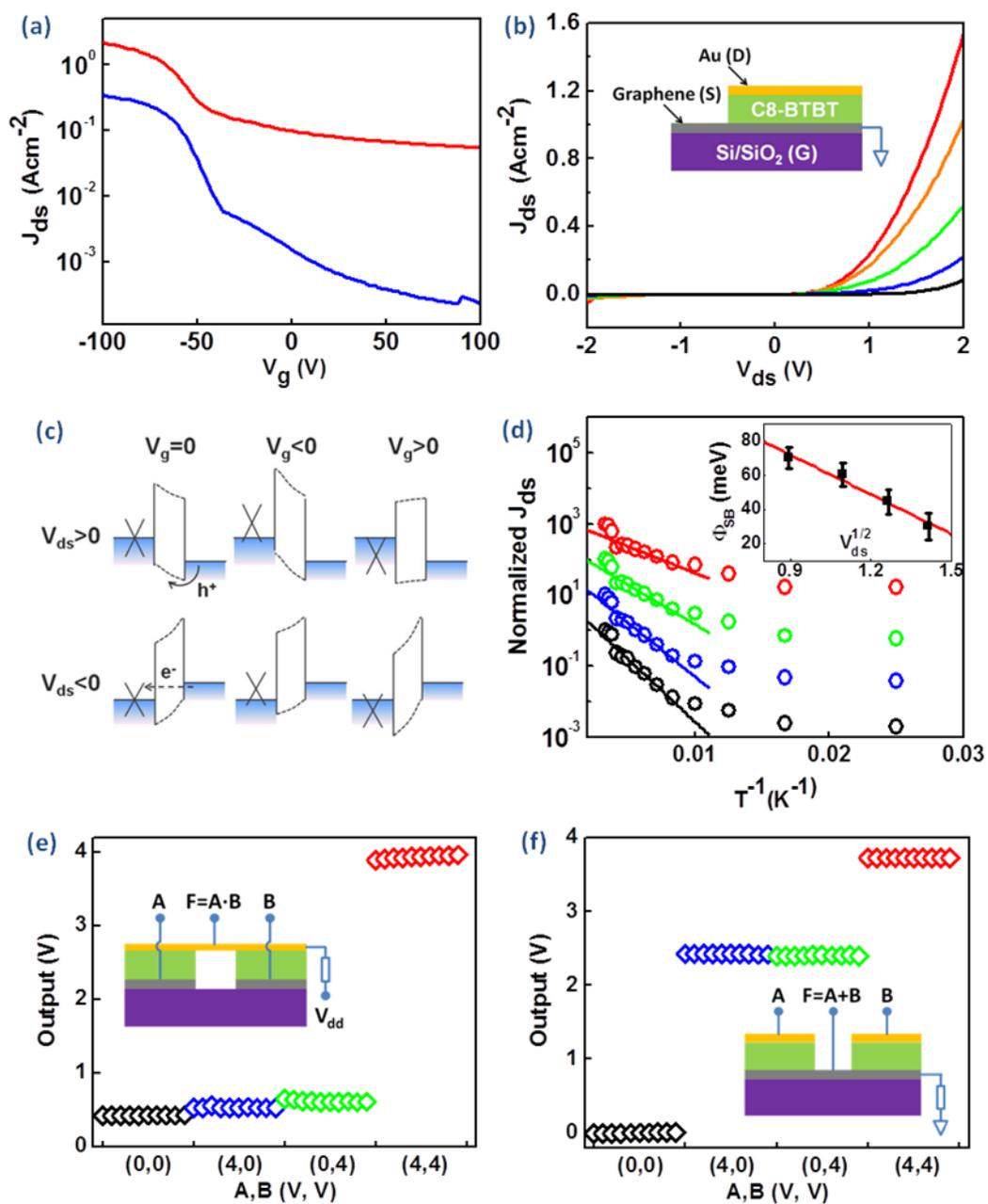

Figure 4:

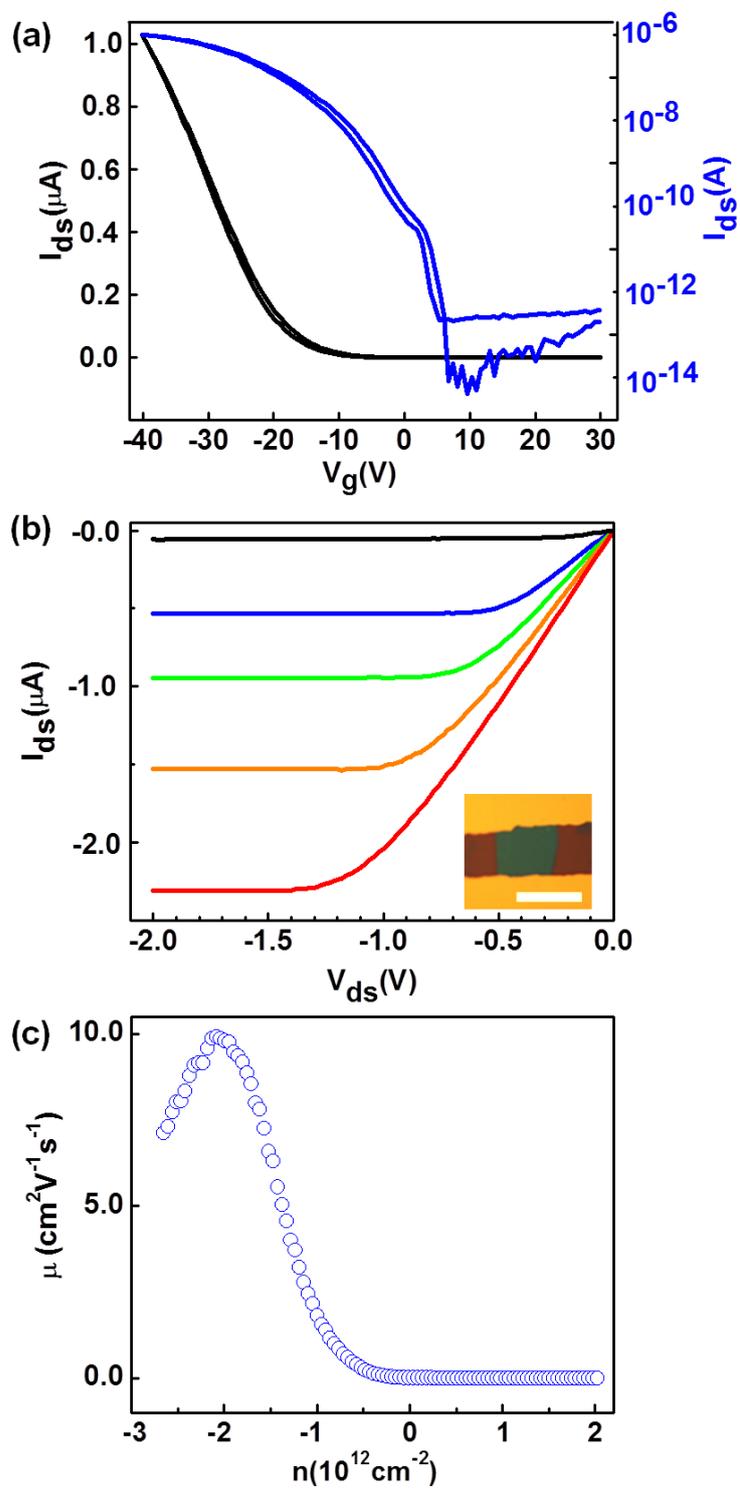

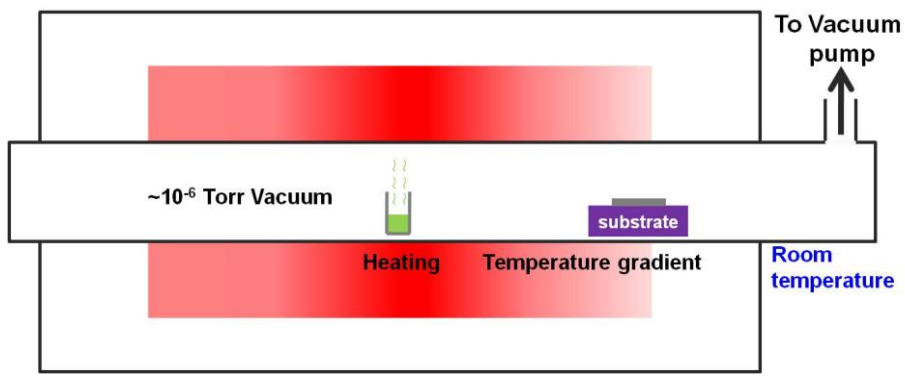

**Supplementary Figure 1 | Experimental setup for crystal growth**. Schematic drawing of the experimental setup for $C_8$-BTBT crystal growth.

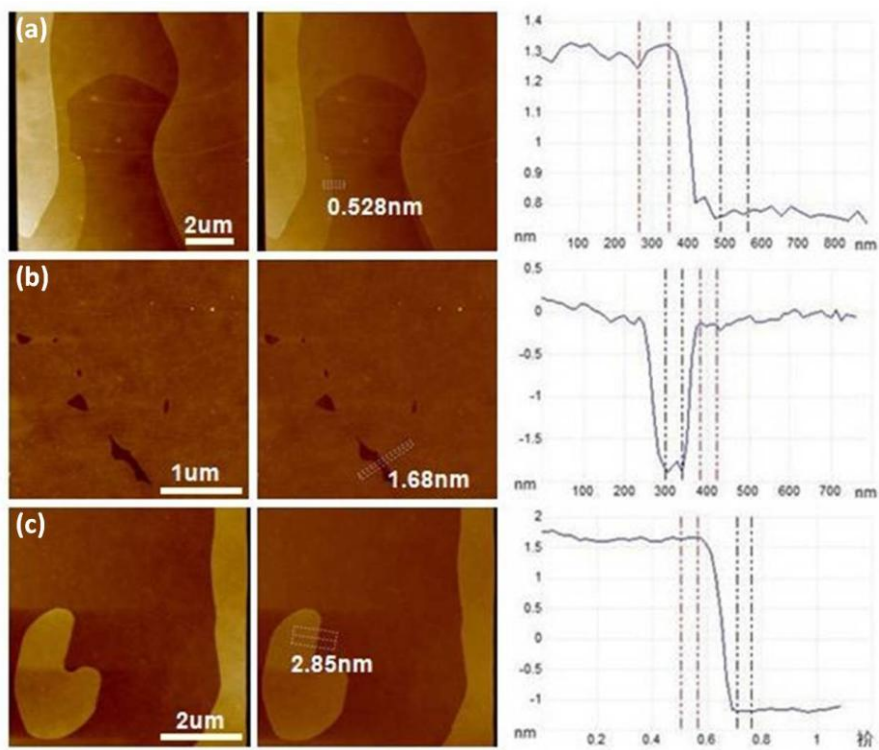

**Supplementary Figure 2 | AFM study of the C$_8$-BTBT crystal growth process on exfoliated graphene and BN.** AFM images of C$_8$-BTBT layers grown on BN, along with the thickness measurement of (a) IL, (b) 1L, (c) 2L. The number of layers is marked on each figure. The thickness of the IL, 1L and 2L sample is ~0.53nm, 1.68nm and 2.85nm, respectively. The heights are similar to those on graphene, suggesting that the molecular packing is the same (Supplementary Figure 24).

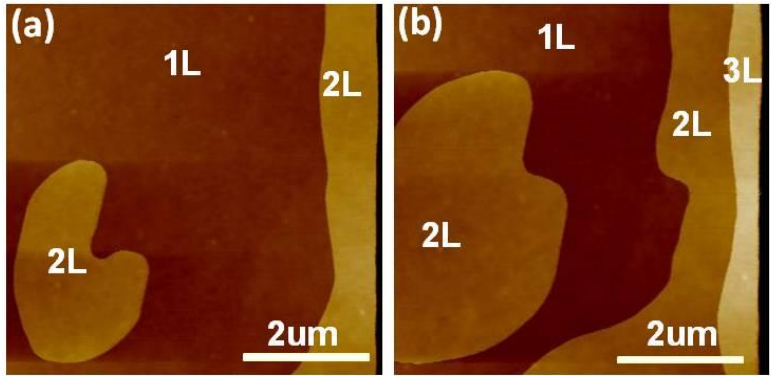

**Supplementary Figure 3 | Sequential AFM images of the growth process of C$_8$-BTBT on BN**. The growth nucleates at certain points and proceeds nearly isotropically in a layer-by-layer manner, similar to the growth on graphene.

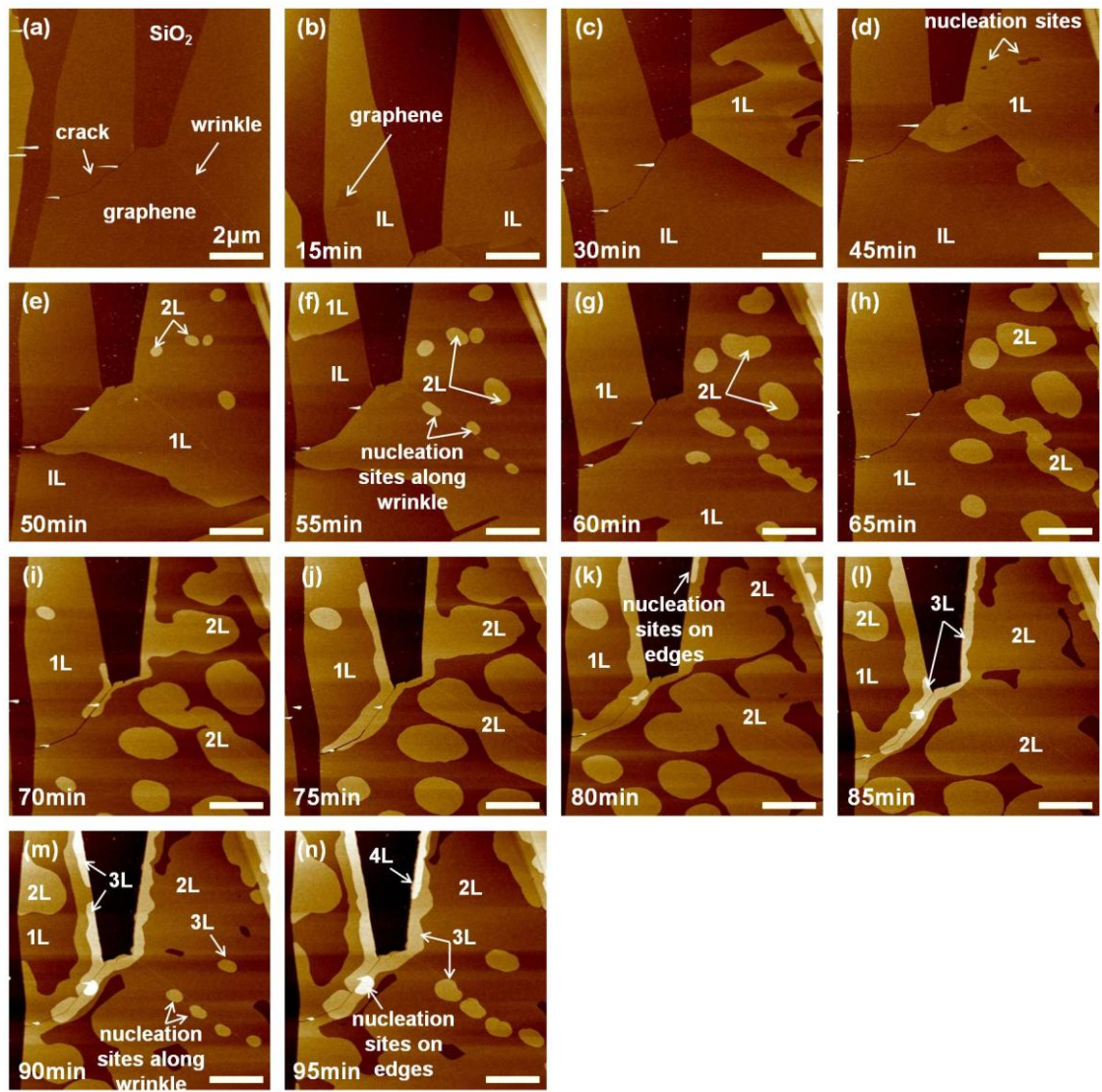

**Supplementary Figure 4 | Sequential AFM snapshots of a graphene sample at different stages during a 95-minute growth.** The scale bars are 2μm. The number of layers, nucleation sites, and the total growth time are marked on each image. (a) is the AFM image of the graphene sample before growth, showing defects such as cracks and wrinkles.

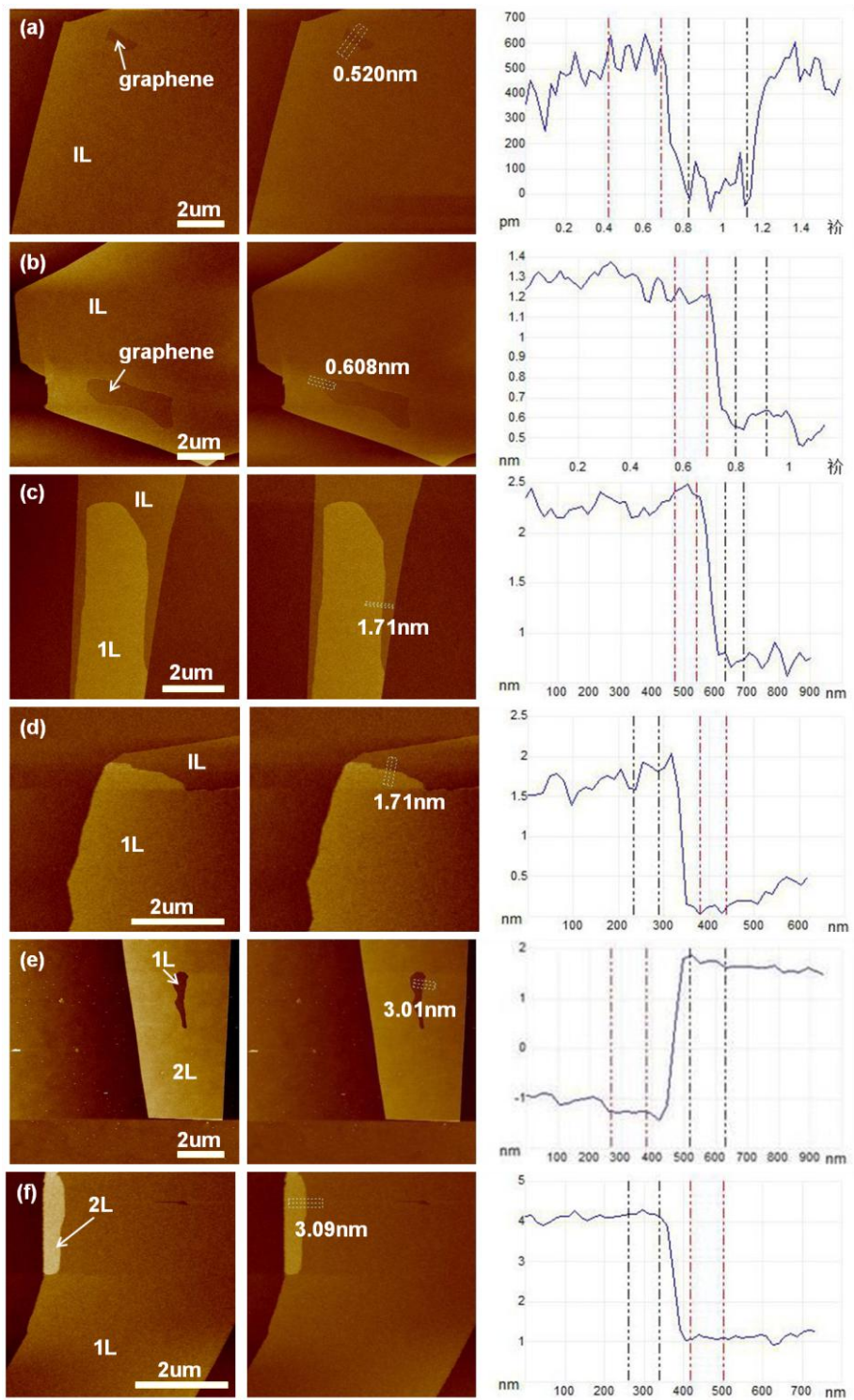

**Supplementary Figure 5 | Thickness measurements of each layer for the first three layers**. AFM images of (a) (b) IL, (c) (d) 1L and (e) (f) 2L $C_8$-BTBT grown on graphene along with the thickness measurements. The number of layers is marked in each image.

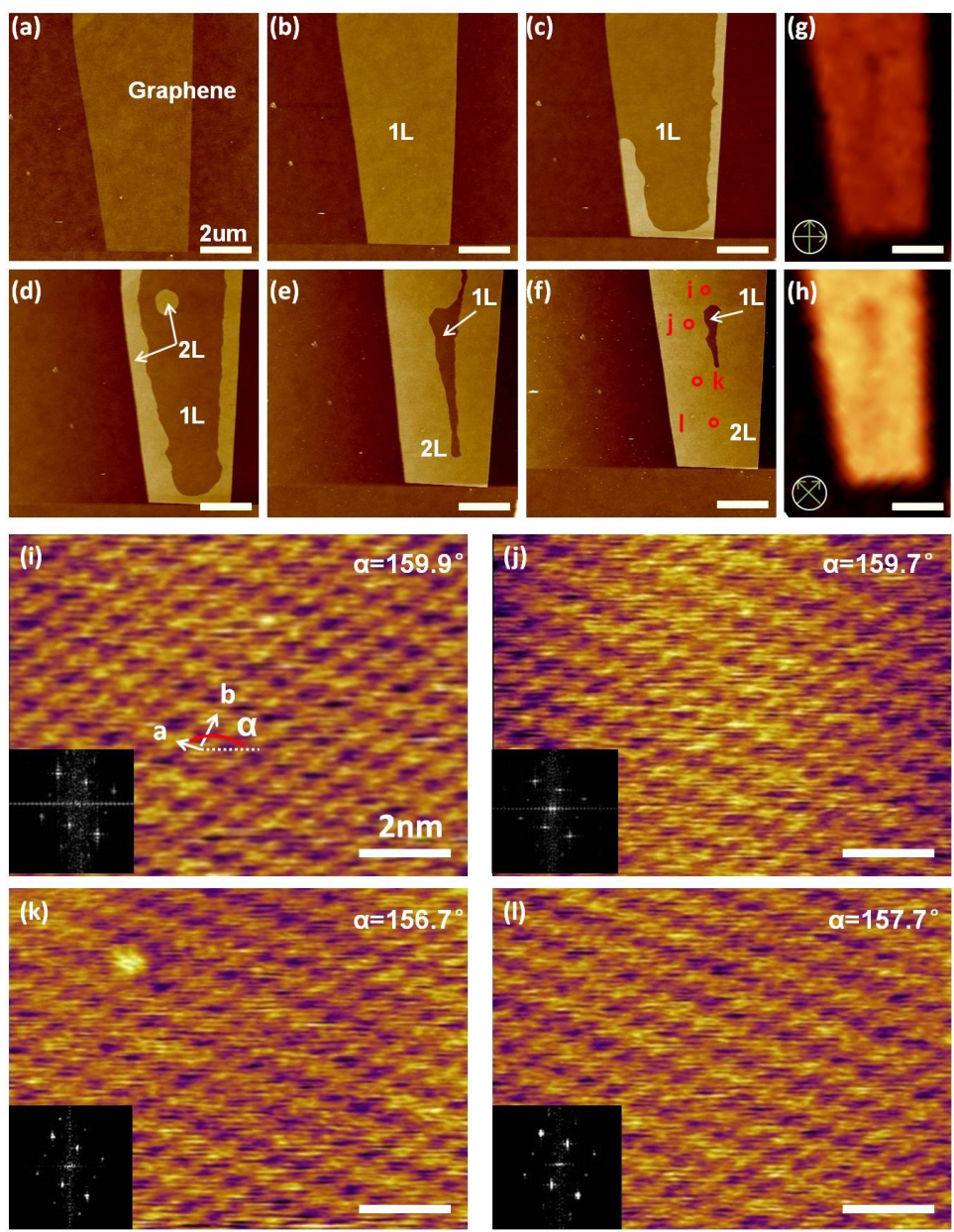

**Supplementary Figure 6 | Sequential AFM images of the growth process of a 2L C$_8$-BTBT on graphene.** (a) is the AFM image of the graphene substrate prior to growth. (b)-(f) are the AFM images taken after growing for 25, 35 55, 100, and 110 minutes, respectively. At stage (b), uniform and complete coverage of 1L is achieved. From (c)-(f), 2L starts to grow from the edges of graphene and proceed uniformly towards the inside. Only one extra nucleation site is observed in (d) during the growth of 2L. (g) and (h) are cross-polarized optical microscopy images of the sample in (f). The uniform color change clearly shows that the whole film is a single crystal. (i)-(l) are high-resolution AFM images taken at the marked positions in (f), where (i) and (j)

are close to one nucleation sites while (k) and (l) are close to the other. The unit cell vectors are marked in (i). $\alpha$ is defined as the angle between **a** and horizontal axis and is marked in every image. The four images show very close lattice orientation, consistent with the cross-polarized microscopy images in (g) and (h). The small difference in lattice orientation could be due to thermal drift of the samples during high-resolution AFM scan. Inset is the Fast Fourier Transform of the AFM image. The scale bars are 2nm.

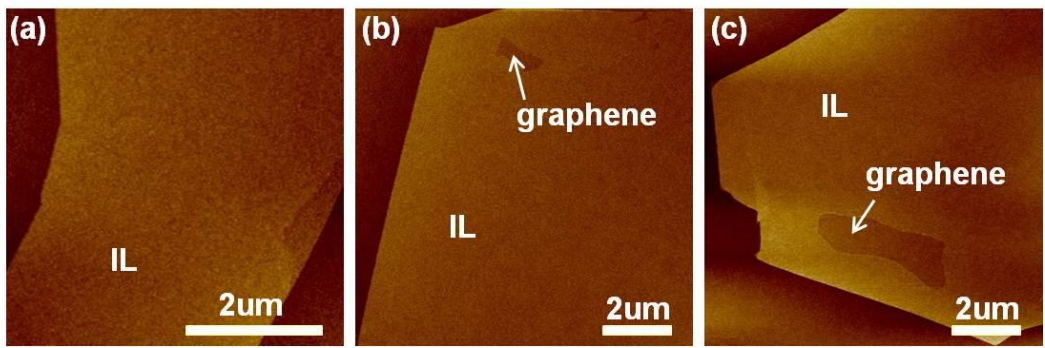

**Supplementary Figure 7 | AFM images of IL C$_8$-BTBT grown on graphene**. The growths are terminated just before the completion of IL to characterize the thickness. Excellent uniformity is achieved in all the samples.

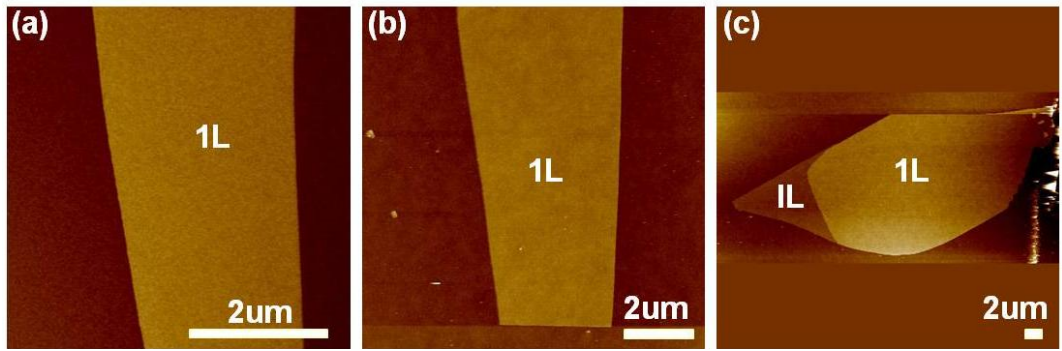

**Supplementary Figure 8 | AFM images of 1L C$_8$-BTBT grown on graphene**. The samples in (a) and (b) are completely covered by 1L of C$_8$-BTBT as determined from the total thickness of the sample. The sample in (c) is partially (but uniformly) covered by 1L of C$_8$-BTBT. The clear boundary between the 1L and IL implies that the growth is epitaxial in nature. Good reproducibility allows us to frequently obtain IL and 1L C$_8$-BTBT (nearly) completely covering graphene by controlling the growth time.

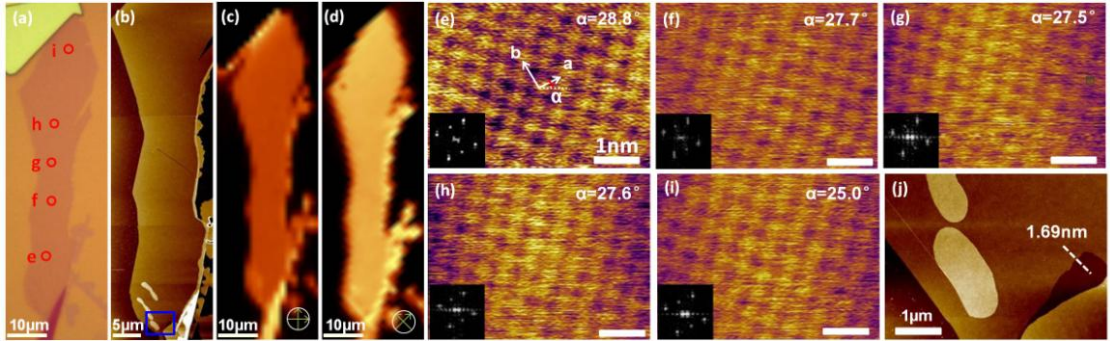

**Supplementary Figure 9 | Characterizations of 1L C$_8$-BTBT grown on exfoliated graphene with size of ~80μm**. (a) Optical microscopy and (b) AFM images of the sample. (c) and (d) are cross-polarized optical microscopy images of the sample. (e)-(i) are the high-resolution AFM images of 1L taken from different spots marked in (a). The unit cell vectors are marked in (e). α is defined as the angle between **a** and horizontal axis and is marked in every image. Inset is the Fast Fourier Transform of the AFM image. The scale bars are 1nm. (j) Zoom-in AFM image from the blue rectangle in (b). The step height of 1.69nm suggests that the whole sample is covered by 1L C$_8$-BTBT. From the AFM and cross-polarized microscopy images, we can confirm single-crystalline 1L C$_8$-BTBT over the entire sample.

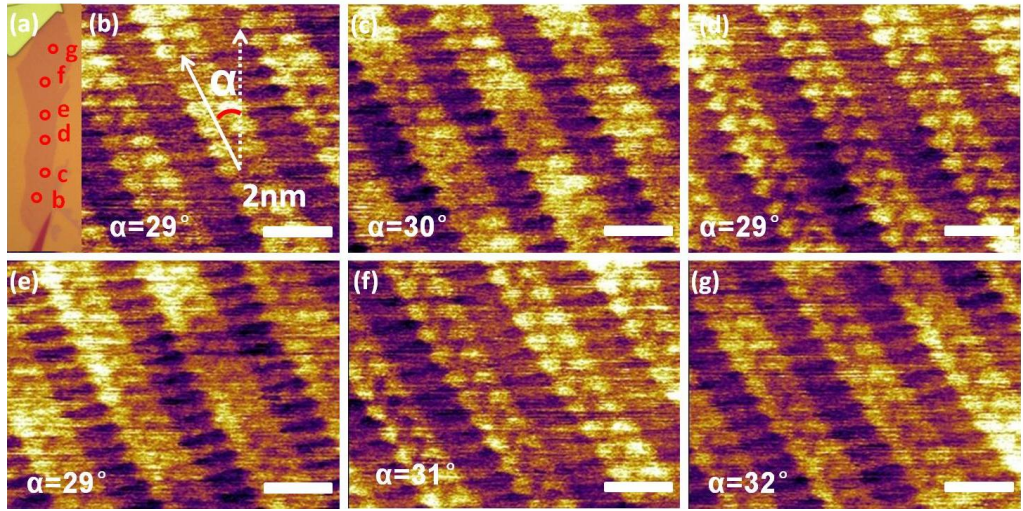

**Supplementary Figure 10 | High-resolution AFM images of the IL $C_8$-BTBT crystal on the same graphene sample in Supplementary Figure 9**. (a) Optical microscopy image of the sample. (b)-(g) High-resolution AFM phase images of the IL $C_8$-BTBT crystal taken at different spots marked in (a). The scale bars are 2nm. Compared to the high-resolution images in Supplementary Figure 9, these images are taken with increased drive amplitude of the cantilever to obtain the information of IL. The crystal orientations match very well for different spots, indicating the single-crystalline nature of IL. The measured periods of two orthogonal directions are 2.72nm and 0.73nm respectively, close to the STM results in the main text.

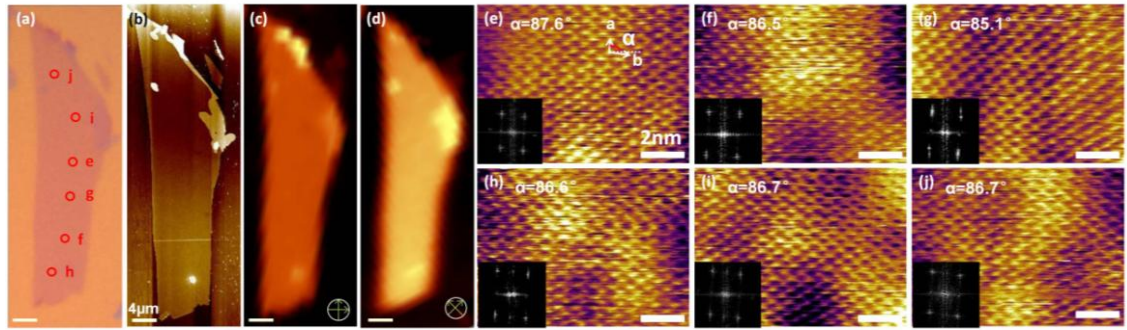

**Supplementary Figure 11 | Characterizations of 1L C$_8$-BTBT grown on exfoliated graphene with size of ~50μm**. (a) Optical microscopy and (b) AFM images of the sample. (c) (d) Cross-polarized optical microscopy images of the sample. The scale bars are 4μm. (e)-(j) are the high-resolution AFM images of 1L taken from different spots in (a). The unit cell vectors are marked in (e). $α$ is defined as the angle between **a** and horizontal axis and is marked in every image. Inset is the Fast Fourier Transform of the AFM image. The scale bars are 2nm. From the AFM and cross-polarized microscopy images, we can confirm single-crystalline 1L C$_8$-BTBT over the entire sample.

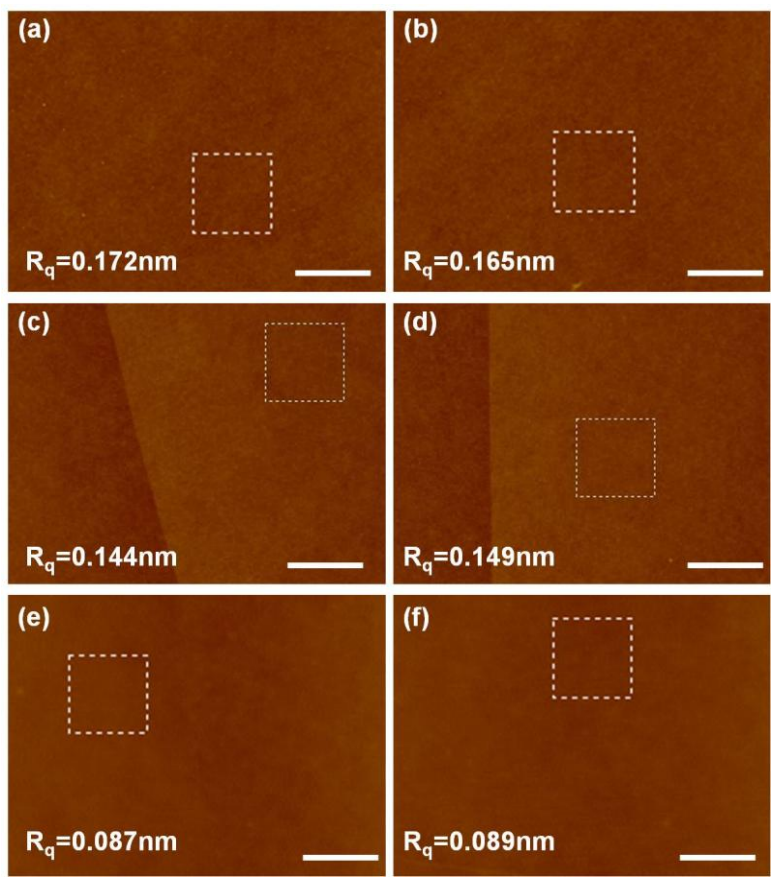

**Supplementary Figure 12 | Root mean square roughness measurements.** Root mean square roughness ($R_q$) analyses of $SiO_2$ substrate (a) and (b), exfoliated monolayer graphene on $SiO_2$ (c) and (d), and BN substrate (e) and (f) used in this work. The $R_q$ values are shown in each figure. The scale bars are 1μm. The $R_q$ of $SiO_2$ and graphene is similar (~0.14-0.17nm) while that of BN is the smallest (~0.09nm), consistent with literature. All the substrates have very small $R_q$, which is important for the high quality epitaxial growth.

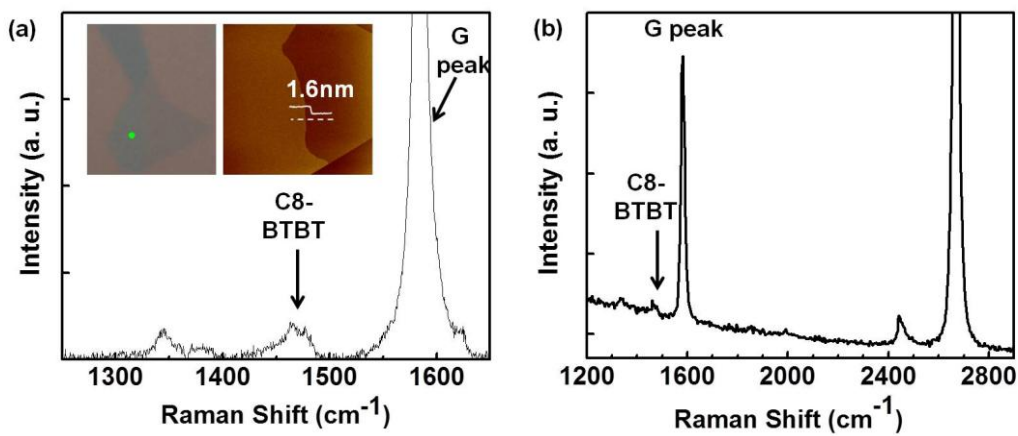

**Supplementary Figure 13 | Raman spectroscopy study of $C_8$-BTBT crystals.** (a) Raman spectrum of $C_8$-BTBT crystals on graphene. The peak near 1470cm$^{-1}$ is from $C_8$-BTBT, consistent with Supplementary Reference 1. Another peak at 1550cm$^{-1}$ overlaps with graphene G peak and is difficult to observe. Left inset shows the optical image of the sample. The green dot marks the position where the spectrum is taken. Right inset shows the AFM image of the $C_8$-BTBT crystal grown on graphene. The 1.6nm step height suggests that the area on the left is 1L. (b) Raman spectrum of the same $C_8$-BTBT crystal in Fig. 1g. The $C_8$-BTBT peak at 1470cm$^{-1}$ is clearly visible, although the intensity is low due to the monolayer thickness of the sample. Fig. 1h shows the Raman mapping of $C_8$-BTBT peak at 1470cm$^{-1}$.

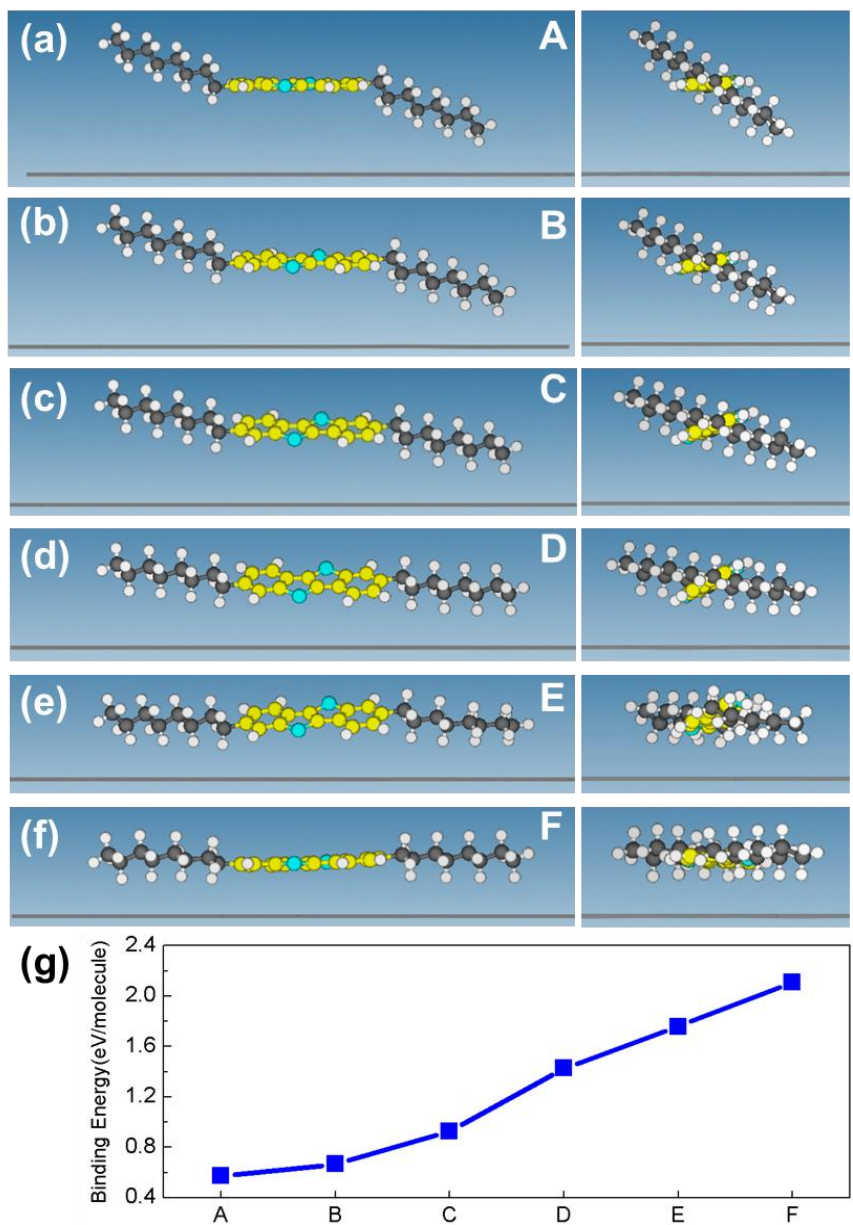

**Supplementary Figure 14 | DFT calculations of the structure of $C_8$-BTBT IL on grapheme.** Different configurations of $C_8$-BTBT molecule adsorbed on graphene (side view) considered in DFT calculations (a-f) and their binding energies (g). Structure A, "stand-up" configuration in which the benzothiophene is parallel to graphene while the carbon chains "stand" up on graphene with the height of 1.04nm. Structures B-D, the $C_8$-BTBT molecule leans gradually on graphene with the angle of 8, 16, and 24 degrees respectively. Structure E, $C_8$-BTBT molecule completely lay down on top of graphene, the alkly chains are almost parallel with the graphene layer. The leaning angle is ~35 degree compared to A. Structure F, from Structure E, the benzothiophene plane is rotated until it is also parallel with the graphene layer. The

grey thick lines in (a)-(f) represent the graphene layer. The binding energy $E_b$ is calculated from the formula *$E_b=E(C_8$-BTBT$)+E(graphene)-E(C_8$-BTBT@graphene)*, where the three terms on the right hand side are the total energy of $C_8$-BTBT molecule, graphene and a single $C_8$-BTBT molecule on graphene, respectively. As expected, Structure F is much more stable than any other configurations because of the strongest *CH-π* interactions and the *π-π* interactions.

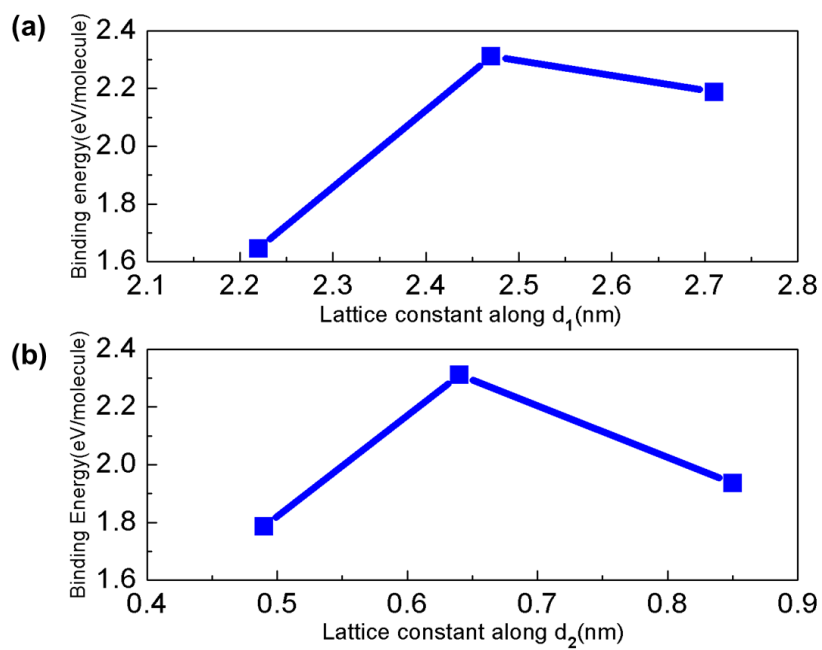

**Supplementary Figure 15 | The binding energy per molecule of C$_8$-BTBT IL crystals on graphene as a function of lattice constant along (a) *d$_1$* direction and (b) *d$_2$* direction.** The structure is the fully relaxed structure shown in Fig. 2b of main text.

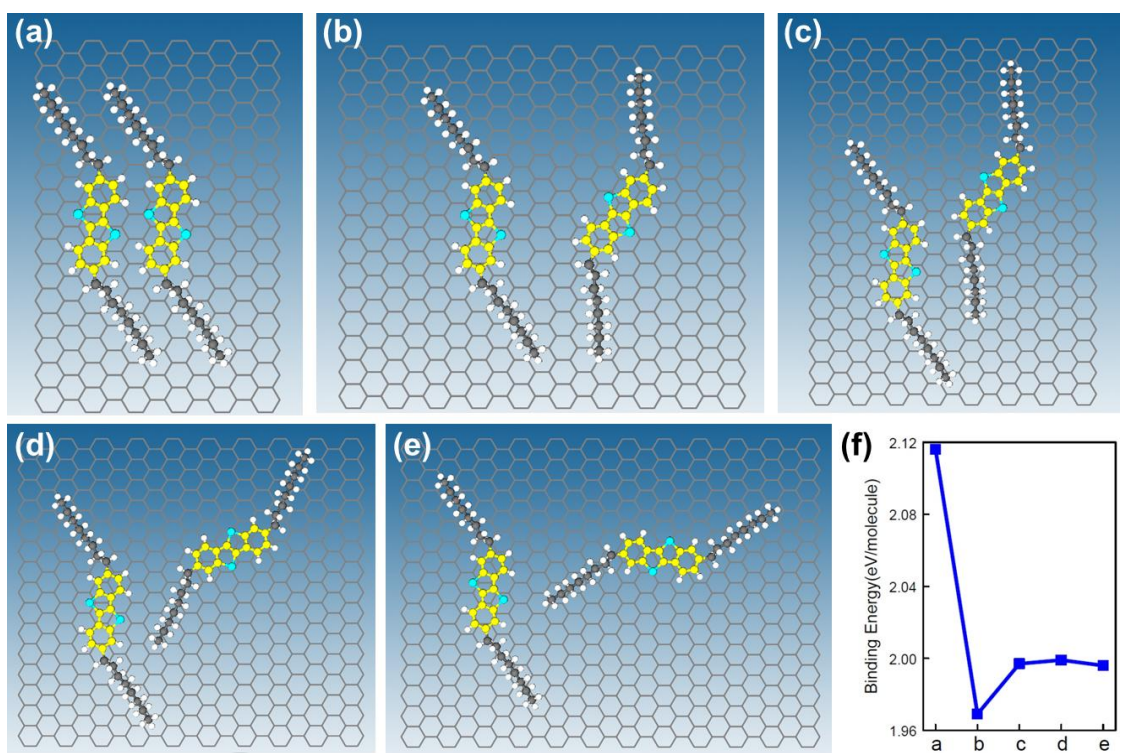

**Supplementary Figure 16 | DFT calculations of the structure of $C_8$-BTBT IL on grapheme.** (a)-(e): Five different configurations of two $C_8$-BTBT molecules on top of graphene. (f) Binding energy per molecule for the different configurations. The configuration in (a) is the most stable one.

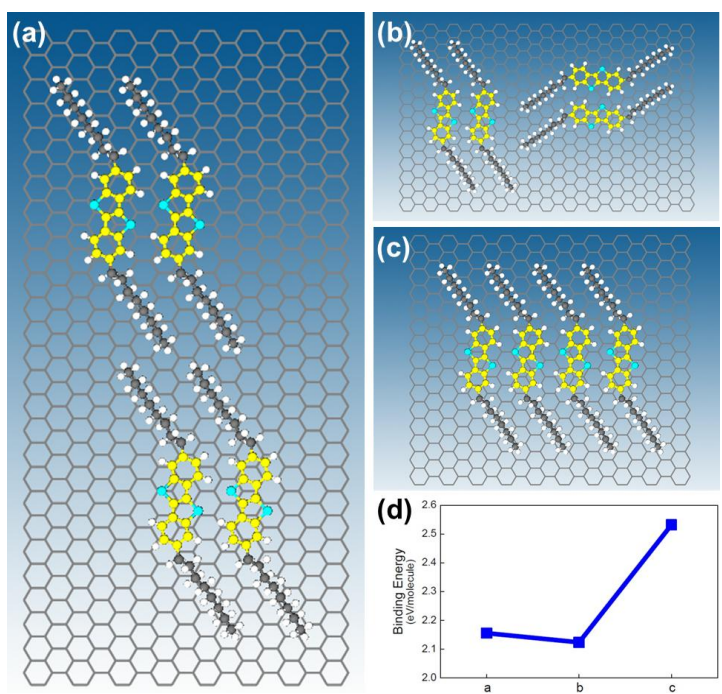

**Supplementary Figure 17 | DFT calculations of the structure of $C_8$-BTBT IL on grapheme**. (a)-(c): Three different configurations of two $C_8$-BTBT dimers on top of graphene. (d) Binding energy per molecule for the different configurations. The configuration in (c) is the most stable one.

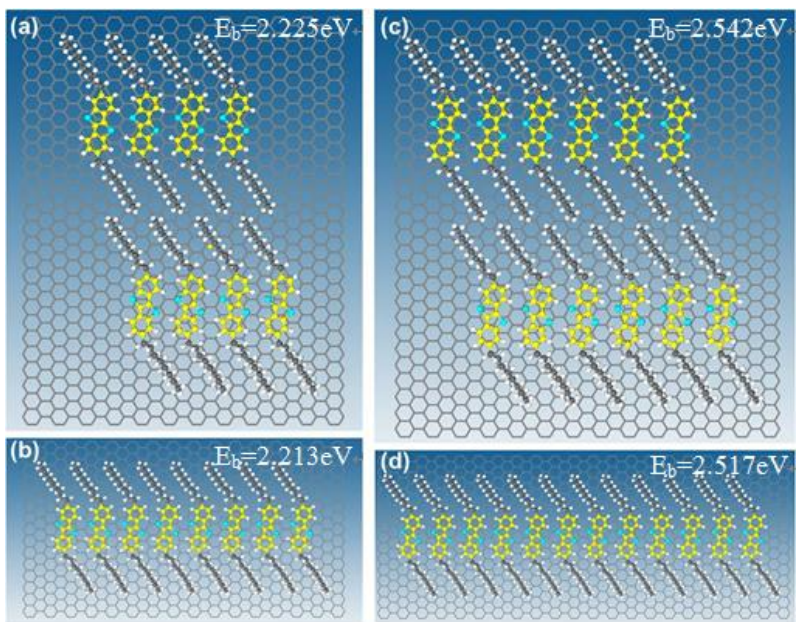

**Supplementary Figure 18 | DFT calculations of the structure of $C_8$-BTBT IL on grapheme**. (a) and (b): Different configurations for 8 $C_8$-BTBT molecules on graphene. (c) and (d): Different configurations for 12 $C_8$-BTBT molecules on graphene. $E_b$ represents binding energies pre molecule . For the 12-molecule cluster, 2D growth in (c) is more stable.

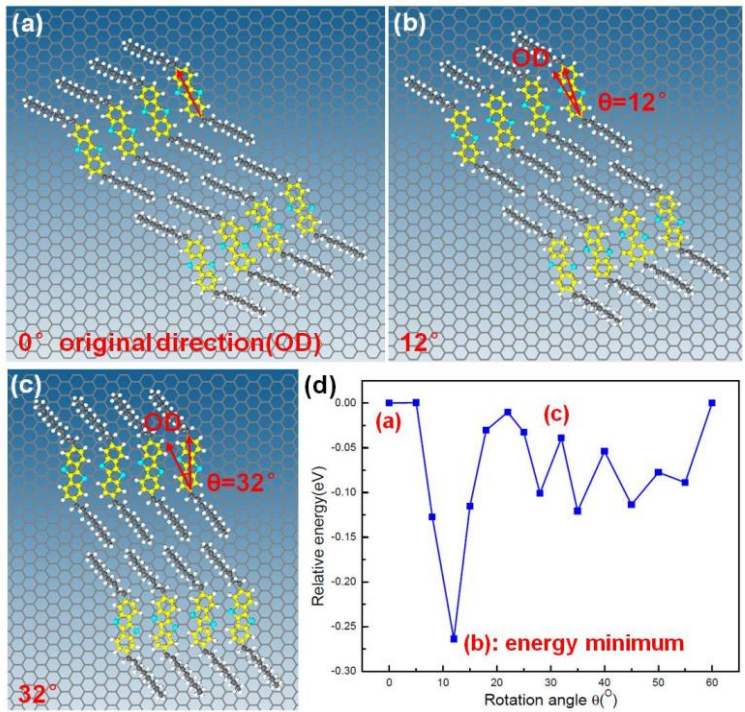

**Supplementary Figure 19 | DFT calculations of the structure of C$_8$-BTBT IL on graphene**. (a)-(c) Different configurations of an 8-C$_8$-BTBT cluster adsorbed on graphene. In (b) and (c), the cluster is rotated clockwise by 12 degrees and 32 degrees respectively with respect to the origin direction in (a), where the benzothiophene align with the armchair direction of graphene layer. (d) Relative energy of the system as a function of the rotation angle between the cluster and graphene. A global energy minimum is observed for the configuration in (b).

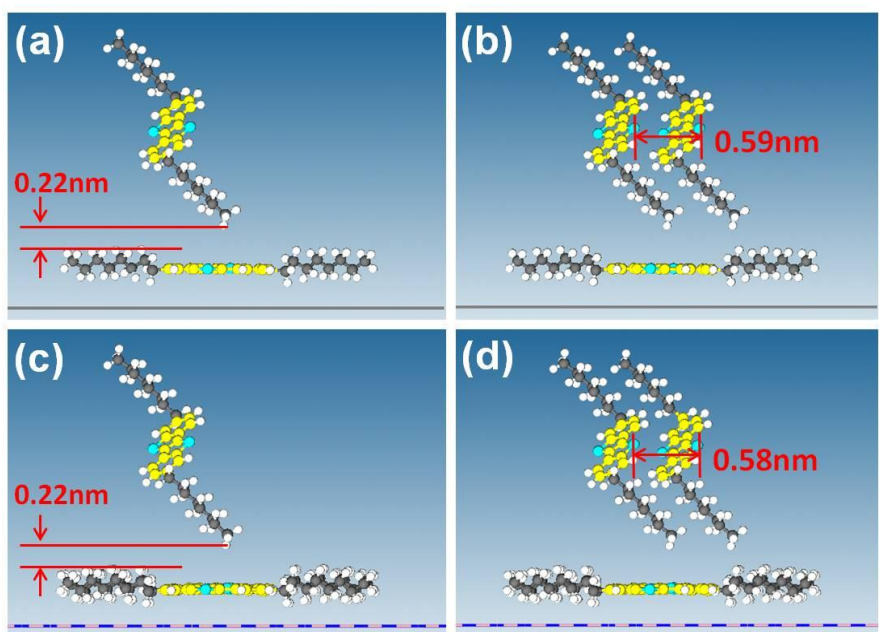

**Supplementary Figure 20 | 1L C$_8$-BTBT molecule adsorbed on IL@graphene and IL@BN.** An 1L C$_8$-BTBT molecule adsorbed on IL@graphene (a) and IL@BN (c). Two 1L C$_8$-BTBT molecules adsorbed on IL@graphene (b) and IL@BN (d). The binding energy is 0.069eV (0.048eV) between the single 1L C$_8$-BTBT molecule and IL@graphene substrate (IL@BN substrate). The binding energy between two 1L C$_8$-BTBT molecules is 0.651eV and 0.704eV for graphene and BN substrate respectively. The fact that interlayer interaction is much weaker than the intralayer interaction suggests the effective decoupling of 1L from the underlying substrate.

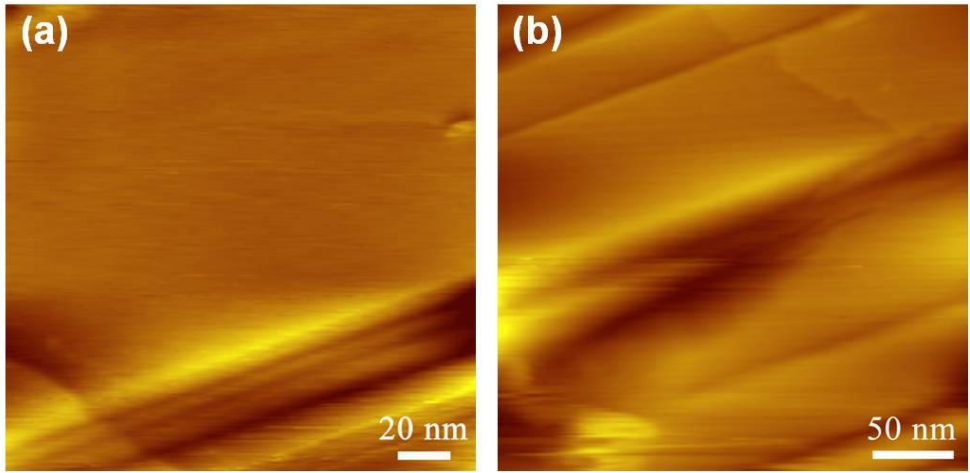

**Supplementary Figure 21 | Large-area STM image of graphene monolayer grown on Cu foils.** (a) is take at $V_{sample}$ = -1.00V and $I$ = 20.6 pA. (b) is taken at $V_{sample}$ = 1.20V and $I$ = 12.7 pA.

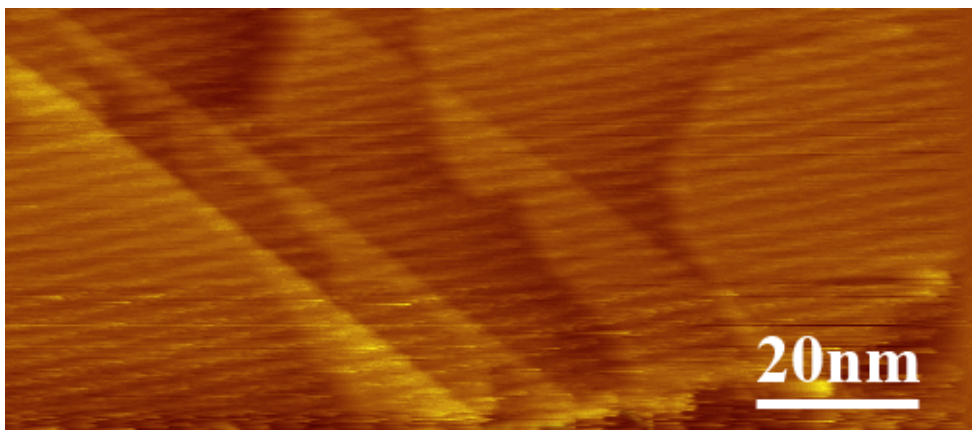

**Supplementary Figure 20 | A topographic STM image of a $C_8$-BTBT IL on graphene/Cu foil recorded at $V_{sample}$ = -1.36 V and $I$ = 4.2 pA**. The periodic structure of the $C_8$-BTBT IL is clearly observed. The single-crystalline structure of the $C_8$-BTBT sheet can maintain over the steps and grain boundaries of the substrate.

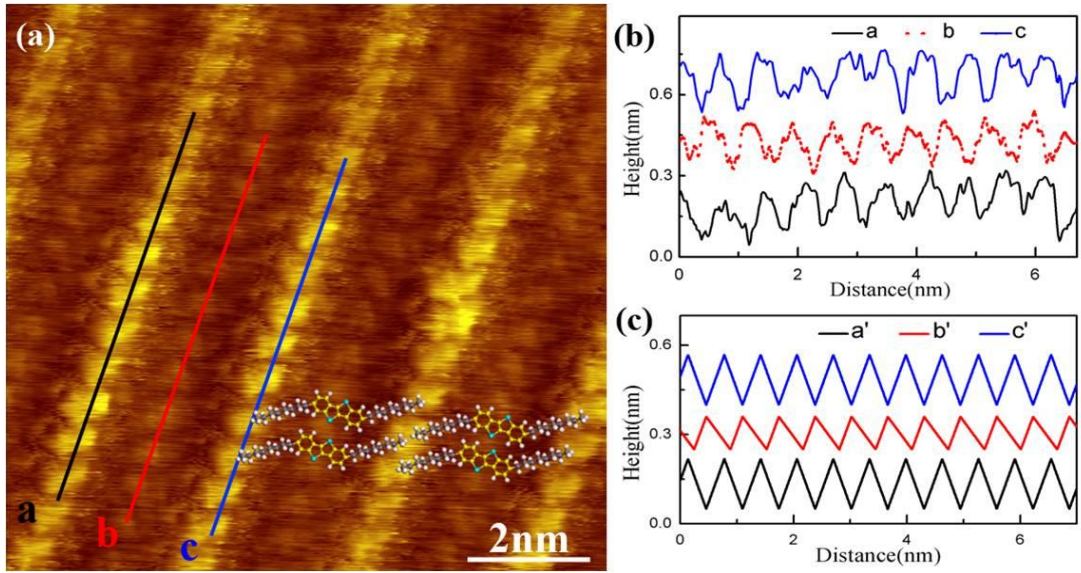

**Supplementary Figure 23** | (a) A typical high-resolution topographic STM image of a $C_8$-BTBT IL on CVD graphene recorded at $V_{sample}$ = -0.91V and $I$ = 11.6 pA. It is easy to observe large-area uniform IL in our samples. The angle between the alkyl chains and the benzothiophene is measured to be $140^o$, which agrees well with the result $\sim 134^o$ obtained from DFT calculation. The atomic structure of the $C_8$-BTBT monolayer is overlaid on the STM image. (b) The section scans along the color lines in panel (a). In (b), we can see a $\pi/2$ phase shift between line a/c and line b. (c) Calculated height profile from the structure in Fig. 2b along the same lines in (b). The phase shift was consistent with experimental results in (b).

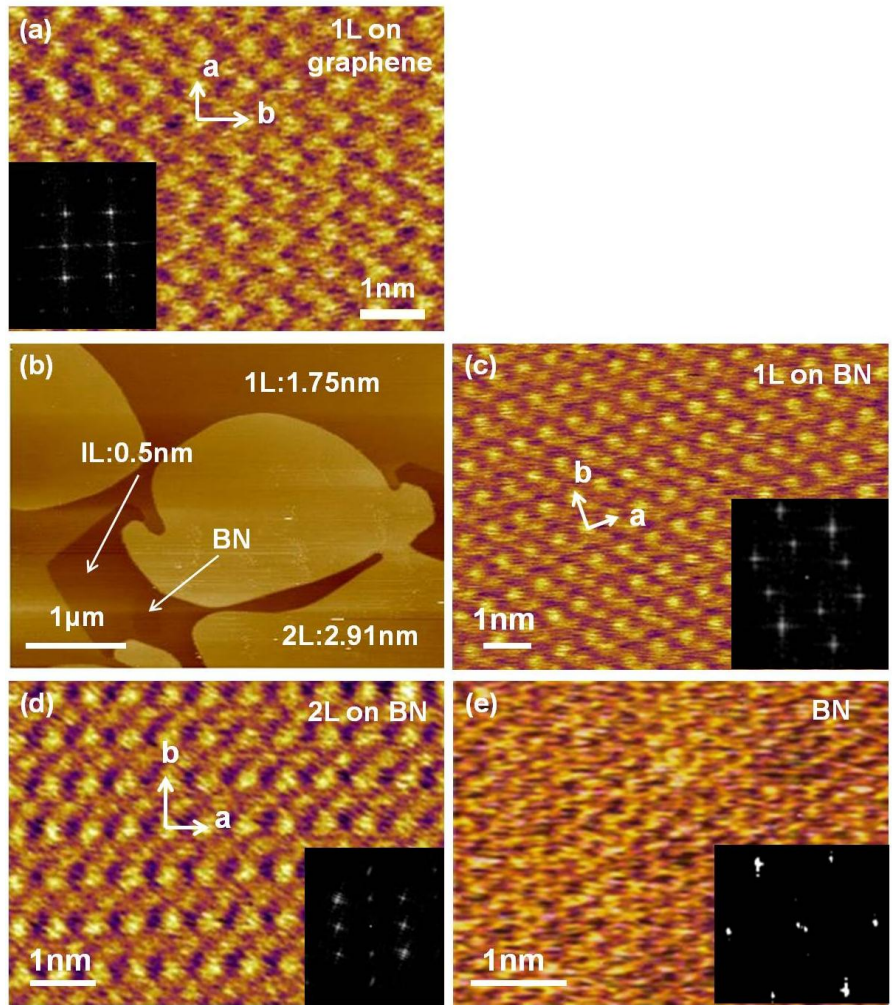

**Supplementary Figure 24 | High resolution AFM studied of $C_8$-BTBT crystals.** (a) High-resolution AFM image of 1L $C_8$-BTBT on graphene. The unit cell is marked. Inset is the Fast Fourier Transform of the AFM image. (b) AFM image of $C_8$-BTBT grown on BN. The number of layers is marked on the image, together with the height of each layer. (c) High-resolution AFM image of 1L $C_8$-BTBT on BN. The unit cell is marked. Inset is the Fast Fourier Transform of the AFM image. (d) High-resolution AFM image of 2L $C_8$-BTBT on BN. The unit cell is marked. Inset is the Fast Fourier Transform of the AFM image. (e) High-resolution AFM image of bare BN substrate. Inset is the Fast Fourier Transform of the AFM image. The lattice is hexagonal as expected, in clear contrast to the $C_8$-BTBT crystals.

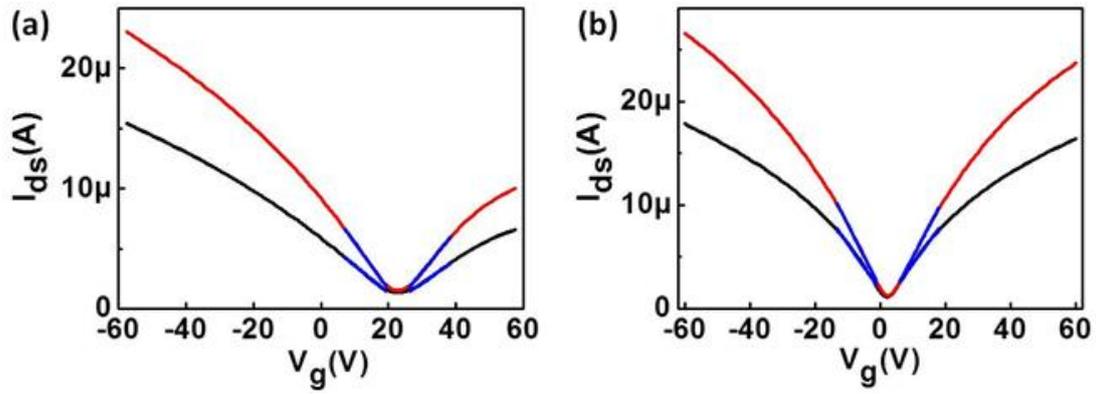

**Supplementary Figure 25 | Transport of graphene before and after deposition of $C_8$-BTBT**. Transfer characteristics of two graphene devices before (black line) and after (red line) deposition of few-layer $C_8$-BTBT. $V_{ds}$=10mV in all cases. Blue lines are theoretical fittings to extract the mobility. The mobility values for electron and hole are listed in the Supplementary Table 1.

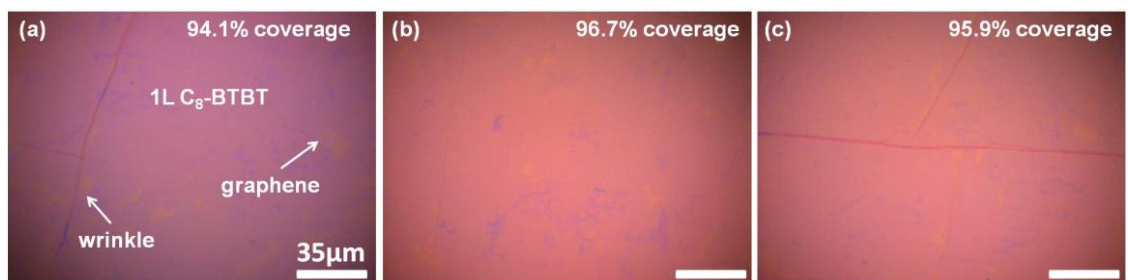

**Supplementary Figure 26 | Optical microscopy images of large-area 1L C$_8$-BTBT grown on CVD graphene**. The coverage of C$_8$-BTBT is marked on every image. The scale bars are 35μm.

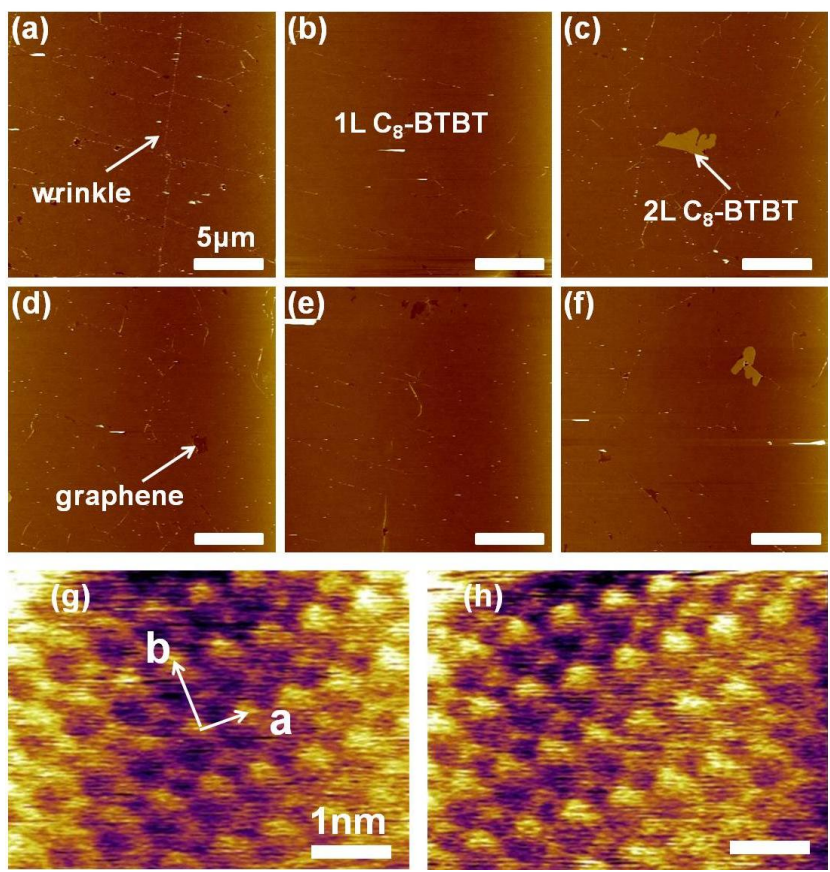

**Supplementary Figure 27 | AFM images of $C_8$-BTBT grown on CVD graphene.** (a)-(f) Large area AFM images of $C_8$-BTBT grown on CVD graphene. The $C_8$-BTBT are predominantly 1L, with small amount of 2L and voids. The scale bars are 5μm. (g) and (h) are high resolution AFM images of 1L $C_8$-BTBT on CVD graphene taken at two different positions. The unit cell vectors are marked in (g). The molecular packing is the same as 1L $C_8$-BTBT on exfoliated graphene. The scale bars are 1nm.

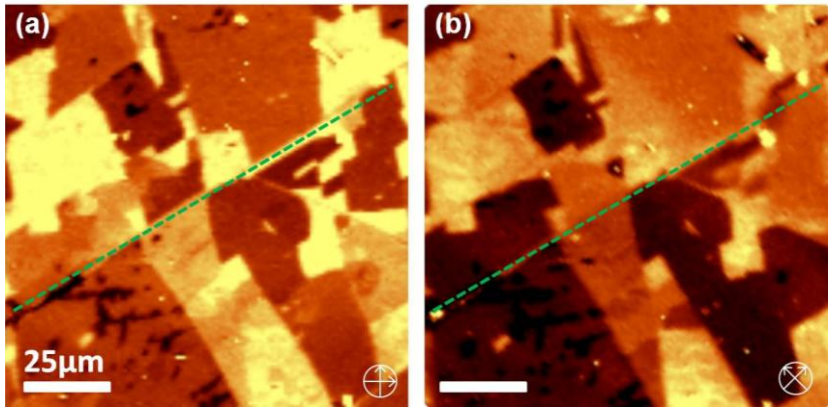

**Supplementary Figure 28 | Cross-polarized optical microscopy images of $C_8$-BTBT film on CVD graphene**. The $C_8$-BTBT is poly-crystalline with average domain size of tens of micrometers. The green dashed lines represent a graphene wrinkle, which strongly affects the growth. Many grain boundaries fall on the wrinkle. The scale bars are 25μm.

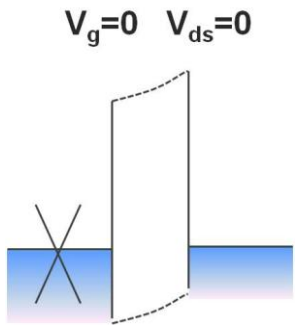

**Supplementary Figure 29 | Energy band diagram of vertical FETs**. Energy band diagram of the vertical FETs under $V_{ds}=0$ and $V_g=0$.

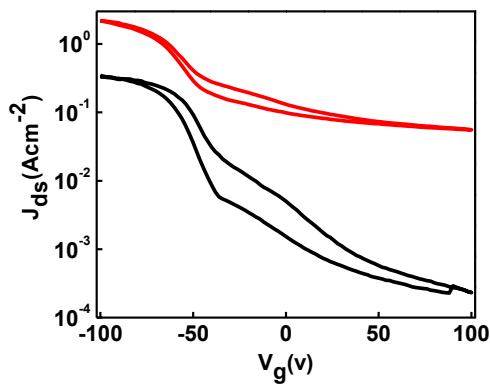

**Supplementary Figure 30 | More data on vertical $C_8$-BTBT OFETs.** Double sweep $J_{ds}$-$V_g$ characteristics of the device in Fig. 3a. From top to bottom, $V_{ds}$=2V and 1V, respectively. We observe small but finite hysteresis in this device. Since the top Au electrodes were transferred on top of $C_8$-BTBT layers in ambient conditions, the hysteresis is likely due to the trapped molecules in the $C_8$-BTBT/Au interface. These molecules can be minimized by shortening the exposure time in ambient, or performing the transfer in a glove box.

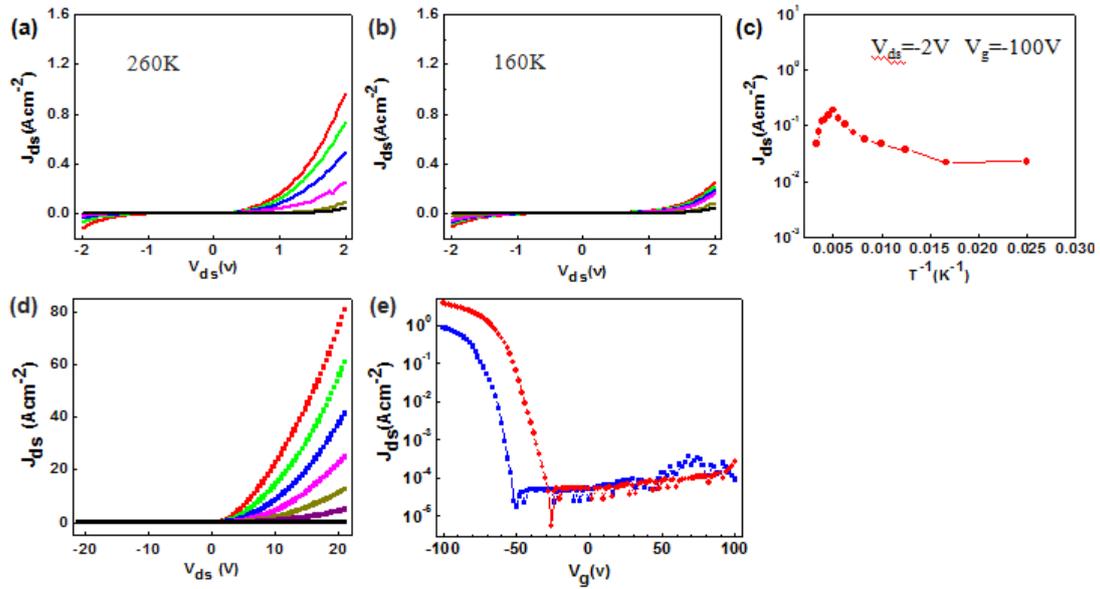

**Supplementary Figure 31 | More data on vertical C$_8$-BTBT OFETs** (a) (b) $J_{ds}$-$V_g$ characteristics of the same device in Fig. 3a at 260K and 160K, respectively. The current is very sensitive to temperature under forward bias but does not change much under reverse bias. From top to bottom, $V_g$=-100V, -90V, -80V, -70V and 0V, respectively. (c) $J_{ds}$ as a function of $1/T$ of the device under $V_{ds}$=-2V and $V_g$=-100V. (d) Room temperature $J_{ds}$-$V_{ds}$ characteristics of another device with 19-layer thick C$_8$-BTBT. From top to bottom, $V_g$=-100V, -90V, -80V, -70V, -60V, -50V and -10V, respectively. (e) Room temperature $J_{ds}$-$V_g$ characteristics of the same device in (d). From top to bottom, $V_{ds}$=2V and 1V, respectively. The $J_{ds}$ at on-state was slightly higher than the thinner device in Fig. 3a, likely due to the difference in contact resistance.

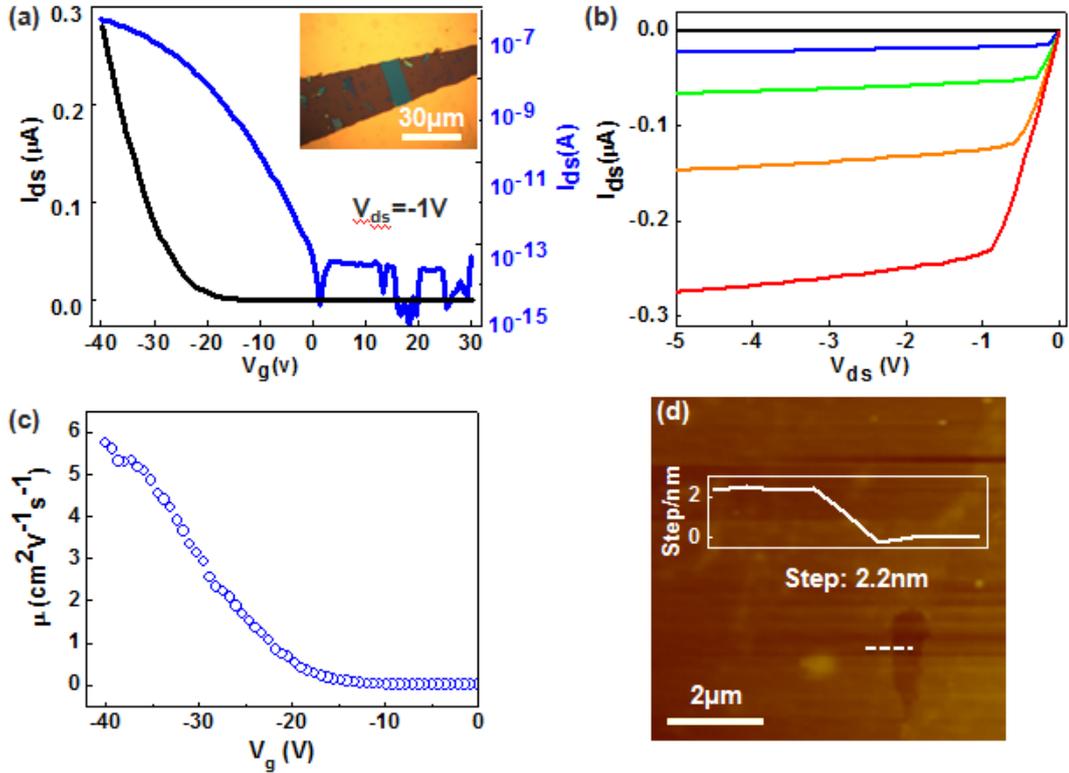

**Supplementary Figure 32 | More data on planar $C_8$-BTBT OFETs**. (a) Room temperature $I_{ds}$-$V_g$ characteristics ($V_{ds}$=-1V) of a planar monolayer $C_8$-BTBT FET on BN. Black and blue lines are drawn in linear and log scales, respectively. Inset shows the optical microscopy image of the device. (b) $I_{ds}$-$V_{ds}$ characteristics of the device. From top to bottom, $V_g$=-10V, -25V, -30V, -35V and -40V, respectively. (c) The extracted μ-$V_g$ relationship at room temperature. The peak mobility is over 5cm$^2$V$^{-1}$s$^{-1}$ for this device. (d) AFM image in the channel of the device after $C_8$-BTBT crystal growth. There exists a small hole with a step height of ~2.2nm (IL+1L), which confirms that the top layer is 1L. The inset shows the height profile along the dashed line.

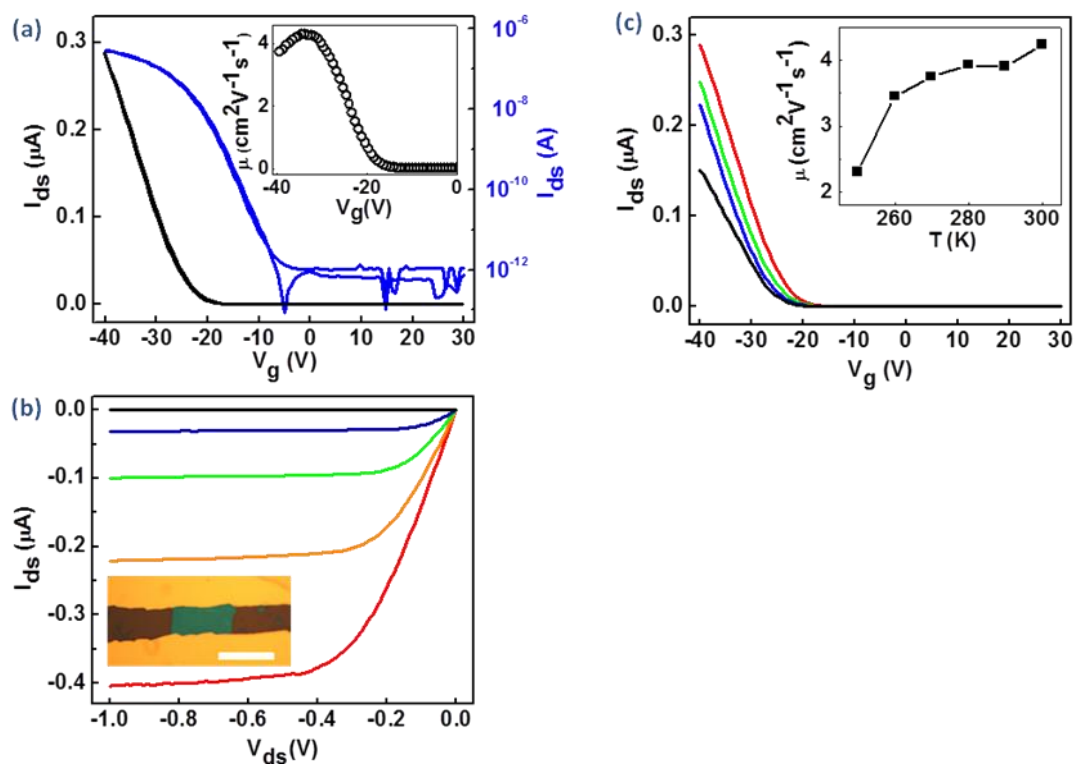

**Supplementary Figure 33 | More data on planar $C_8$-BTBT OFETs.** (a) Room temperature double-sweep $I_{ds}$-$V_g$ characteristics ($V_{ds}$=-0.2V), with little hysteresis. Black and blue lines are drawn in linear and log scales respectively. Inset shows the extracted $\mu$-$V_g$ relationship, the peak mobility is 4.3cm$^2$V$^{-1}$s$^{-1}$. (b) $I_{ds}$-$V_{ds}$ characteristics of the device in (a). From top to bottom, $V_g$=-10V, -25V, -30V, -35V and -40V, respectively. Inset shows the optical microscopy image of the device. Scale bar, 30μm. (c) Variable-temperature $I_{ds}$-$V_g$ characteristics ($V_{ds}$=-0.2V) of the device in (a) at 300K, 290K, 270K and 250K, respectively (from top to bottom). Inset shows $\mu$ as a function of temperature.

|  | $\mu_e$ before deposition | $\mu_e$ after deposition | $\mu_h$ before deposition | $\mu_h$ after deposition |
| --- | --- | --- | --- | --- |
| Device a | 5714 | 10125 | 6809 | 11546 |
| Device b | 12624 | 17683 | 13216 | 18184 |

**Supplementary Table 1 | The mobility values for electron and hole**. The high mobility values after deposition strongly suggest the weak coupling between graphene and $C_8$-BTBT layers.

**Supplementary note 1: Molecular structure of $C_8$-BTBT molecules on graphene.**

In order to understand the growth process of IL, we first need to find the most stable configuration of $C_8$-BTBT dimers as the building block. In Supplementary Figure 16 we calculate the energy of five different configurations, and find that the most stable configuration appears when the two molecules are closely packed. This is expected because the interaction between the two molecules is strongest in this configuration.

Next, using the most stable dimer as the building block, we can build the quadrumer. Supplementary Figure 17 shows three different configurations, and the most stable one is (c) where the interaction between the dimers is the strongest. At this stage, 1D growth is still favorable.

Similarly, we can build larger clusters with the building blocks. However, as we increased the size of the cluster to some point, 1D growth starts to become less favorable because interaction only occurs between one pair of $C_8$-BTBT molecules. At this point, 2D growth happens. In Supplementary Figure 18 we calculate the clusters with 8 and 12 $C_8$-BTBT molecules. For the 8-molecule cluster, 1D and 2D configuration have very similar binding energies (2D growth is slightly more stable). For the 12-molecule cluster, the preference of 2D growth becomes more evident. Therefore, we conclude that the 2D growth of IL $C_8$-BTBT molecules is a natural result of the interactions between the molecules.

**Supplementary note 2: large-area single crystals of $C_8$-BTBT epitaxially grow on graphene**

To understand why large-area single crystals of $C_8$-BTBT can epitaxially grow on graphene, we further calculate the total energy of a $C_8$-BTBT cluster and graphene as a function of the rotation angle between them (Supplementary Figure 19). We find that there exists a global energy minimum, where the IL $C_8$-BTBT is the most stable. This explains why IL $C_8$-BTBT can form a large single crystal on graphene (Supplementary Figure 10), even with multiple nucleation sites. However, we note that the energy difference is on the order of 0.2eV, which is not very large. So it is still possible to have multiple domains on the same exfoliated graphene, as shown in Fig.

1i of main text.

**Supplementary note 3: STM study of $C_8$-BTBT crystals on graphene**

In our STM experiment, we use monolayer CVD graphene grown on Cu foil as the substrate. Before the growth of the $C_8$-BTBT films, the graphene on Cu foil was studied carefully by STM. Supplementary Figure 21 shows two typical large-area STM images recorded at different positions of the as-grown sample. The sample is overall very flat with small corrugated areas and with some steps on the copper surface. Our STM studies indicate that the graphene sheet can cross over the copper steps and maintain a continuous structure. Since high resolution STM scans are usually recorded in small areas like 10nm × 10nm, the substrate roughness is very small and does not affect the STM measurement. Supplementary Figure 22 shows a large-area STM image of the IL $C_8$-BTBT film on CVD graphene. The periodic structure in the $C_8$-BTBT film can clearly be observed. The $C_8$-BTBT over-layer can maintain a continuous single-crystalline structure over the grain boundaries and steps of the graphene surface. This result also implies that the interaction between graphene and the $C_8$-BTBT is very weak, consistent with the vdW epitaxy.

**Supplementary note 4: $C_8$-BTBT crystal growth on CVD graphene**

The growth on CVD graphene has two major challenges compared to exfoliated graphene. 1. Our CVD graphene samples are poly-crystalline with average domain size on the order of tens of micrometers. 2. After transfer, there are unavoidable wrinkles, cracks and small amount of PMMA residue on graphene samples, which strongly affect the growth of $C_8$-BTBT crystals (see Fig. 1 of main text and related discussions). Despite these challenges, we are able to grow large-area $C_8$-BTBT films on CVD graphene, with over 90% coverage. Supplementary Figure 26 and 27 show the optical and AFM images of typical $C_8$-BTBT films on CVD graphene. The $C_8$-BTBT films are predominantly 1L, with small areas of voids and 2L.

To further characterize the $C_8$-BTBT films on CVD graphene, we carry out cross-polarized microscopy (Supplementary Figure 28). The films are poly-crystalline, with similar domain size as the graphene substrate. This is not surprising considering the epitaxial nature of the growth. Therefore, we attribute the poly-crystalline nature mainly to the underlying graphene substrate. But other impurities such as winkles and cracks also affect the growth and are expect to make the $C_8$-BTBT domain even smaller (Supplementary Figure 28).

**Supplementary References:**

1. Li, Y., Liu, C., Lee, M. V., Xu,Y., Shi, Y., Tsukagoshi, K. In situ purification to eliminate the influence of impurities in solution-processed organic crystals for transistor arrays. *J. Mater. Chem. C* **1**, 1352-1358 (2013).
2. Northrup, J. E., Tiago, M. L. & Louie, S. G. Surface energetics of and growth of pentacene. *Phys. Rev. B* **66**, 121404(R) (2002).